\documentclass[letterpaper,twocolumn,english,aps,prl,groupedaddress,superscriptaddress,showpacs,amsfonts]{revtex4}
\usepackage{ae}
\usepackage{aecompl}
\usepackage[T1]{fontenc}
\usepackage[latin1]{inputenc}
\usepackage{float}
\usepackage{amsmath}
\usepackage{graphicx}
\usepackage{amssymb}

\makeatletter
\usepackage{babel}
\makeatother
\begin{document}

\title{Magnetic Interaction between Surface Engineered Rare-earth Atomic Spins}

\author{Chiung-Yuan Lin}

\affiliation{Department of Electronics Engineering, National Chiao
Tung University, Hsinchu, Taiwan}

\author{Jheng-Lian Li}

\affiliation{Department of Electronics Engineering, National Chiao
Tung University, Hsinchu, Taiwan}
\affiliation{Research Center for Biomedical Devices, Taipei Medical University, Taipei, Taiwan}

\author{Yao-Hsien Hsieh}

\affiliation{Department of Electronics Engineering, National Chiao
Tung University, Hsinchu, Taiwan}

\author{B.~A.~Jones}

\affiliation{IBM Almaden Research Center, San Jose, CA 95120-6099, USA}

\date{\today}

\begin{abstract}
We report the ab initio study of rare-earth adatoms (Gd) on an insulating surface.
This surface is of interest because of previous studies by scanning tunneling
microscopy showing spin excitations of transition metal adatoms
\cite{MnScience}.
The present work is the first study of rare-earth spin-coupled adatoms, as well as
the geometry effect of spin coupling, and the underlying mechanism of ferromagnetic coupling.
The exchange coupling between Gd atoms on the surface is calculated
to be antiferromagnetic in a linear geometry and ferromagnetic in a diagonal geometry,
by considering their collinear spins and using the PBE+U exchange correlation.
We also find the Gd dimers in these two geometries are similar to the nearest-neighbor (NN) and
the next-NN Gd atoms in GdN bulk.
We analyze how much direct exchange, superexchange, and RKKY interactons
contribute to the exchange coupling for both geometries
by additional first-principles calculations of related model systems.
\end{abstract}

\maketitle

Understanding the spin coupling at the nanoscale is important in scaling down magnetic devices,
such as spintronics and quantum computing devices \cite{qbit}.
During the past years, it has been demonstrated that the scanning tunnelling microscope (STM)
is a powerful tool to build and control individual nano magnetic structures,
and has a great potential to being applied to the construction of nanoscale magnetic devices.
Previous studies of engineering individual magnetic atoms on surfaces include
the antiferromagnetism of Mn chains ($1\sim10$ atoms) \cite{MnScience},
the anisotropy of a single Fe atom \cite{FeScience} and a Fe-Cu dimer \cite{FeCuScience},
Kondo effects under magnetic environments \cite{CoNaturePhysics,CoPRL},
and a bistable atomic-scale Fe antiferromagnet that demonstrates
the feasibility of dense nonvolatile storage of information
at low temperature \cite{FechainScience},
where all above magnetic systems
are placed on top of a copper nitride island
that serves as an insulating monolayer on a Cu(100) substrate.

Before physicists started using the STM to manipulate and couple magnetic atoms together,
chemists have decades of history in synthesizing numerous species of molecules
that carry giant spins, well-known as molecular magnets \cite{MM}.
The STM-engineered spins were found to form a
surface molecular network with great similarity to molecular magnets.
While the major attention to the field of molecular magnets is mainly focused
on those consisting of transition-metal magnetic atoms, very few studies are
devoted to the rare-earth-based molecular magnets.
Moreover, only magnetic anisotropy is studied for the rare-earth-based molecular magnets \cite{ReSMM},
but not the interatomic spin coupling within such molecules.
A similar situation also happens to the STM-engineered spins;
experimentalists have not tried to place rare earth atoms on the CuN/Cu(100) surfaces
to see how their spins couple to each other.

Rare earth atoms have their magnetism primarily contributed from the $f$ orbitals,
and may behave quite differently from those well-studied transition atoms
when being placed on the surface.
Among the lanthanoid series,
a free Gd atom has a half-filled $4f$ shell such that it carries an atomic spin
much larger than the transition atoms, and at the same time has an $f$ shell of $L=0$
that is expected to exhibit quite small magnetic anisotropy on surfaces
due to small spin-orbit interaction.
The spin excitations of spin coupling and
magnetic anisotropy can both contribute to the STM inelastic tunnelling spectroscopy.
If both types of excitations exist at the same energy scale,
it makes the inelastic tunnelling spectra difficult to be analyzed.
On the other hand, with low anisotropy one would expect that Gd atoms on the CuN surface yield clean
inelastic tunnelling spectra mainly from the interatomic coupling of their spins.
Such an advantage would benefit future experimental studies following this first-principles investigation.

With the goal of making nano magnets by atom manipulation,
one needs to couple atomic spins ferromagnetically on such a surface.
Previous first-principles calculations show
the antiferromagnetism originated from superexchange interactions along a linear path \cite{MnDFT}.
The Gd atoms also serve as a potential candidate for ferromagnetic atomic-spin dimers, as we describe below.

In this work, we perform first-principles calculations of Gd adatoms on the CuN surface.
In some ways, the Gd atoms are similar to the previously-studied Mn atoms \cite{FeScience,MnDFT}
when being deposited on the Cu sites of the CuN surface, i.e.,
the Gd's nearby N atoms break bounds with their neighboring Cu and
form a "quasi" molecular structure from the surface.
However, the local structures of the Gd atoms on the CuN surface have
a well-studied reference system, the GdN bulk.
We build two different geometries of the Gd dimers on the surface:
one has Gd atoms along the same N row, and the other along two perpendicular N rows.
The two geometries mimic the coupling paths of the NN and next-NN
Gd atoms of the GdN bulk, where the two paths in bulk have
ferromagnetic and antiferromagnetic couplings, respectively.
We calculate the exchange couplings $J$ of two arrangements of Gd$_{2}$/CuN
using first-principles PBE+U, and expect that one of the two types of surface Gd dimers
will exhibit ferromagnetism and the other antiferromagnetism.

\begin{figure}
\begin{center}
\includegraphics[keepaspectratio,trim=0cm 0cm 0cm 0cm,clip,height=5cm]{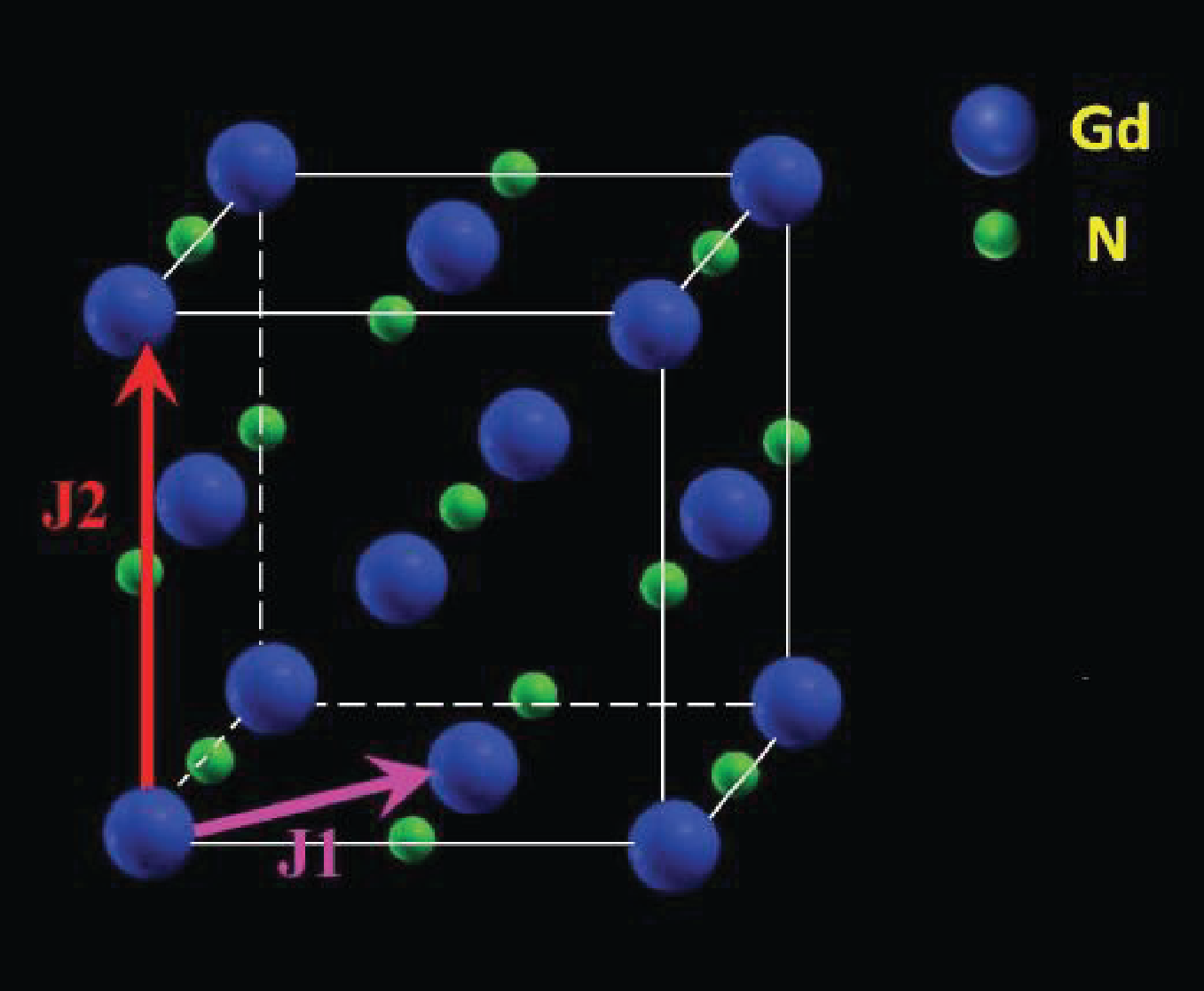}
\end{center}
\caption{\label{GdN}
The unit cell of a GdN bulk. The Gd-to-Gd arrows indicate the NN
(diagonal, purple)
and next-NN (linear, red) couplings.}
\end{figure}

\begin{figure}
\begin{center}
\includegraphics[keepaspectratio,clip,height= 3.7cm,trim=0cm 0cm 0cm 0cm]{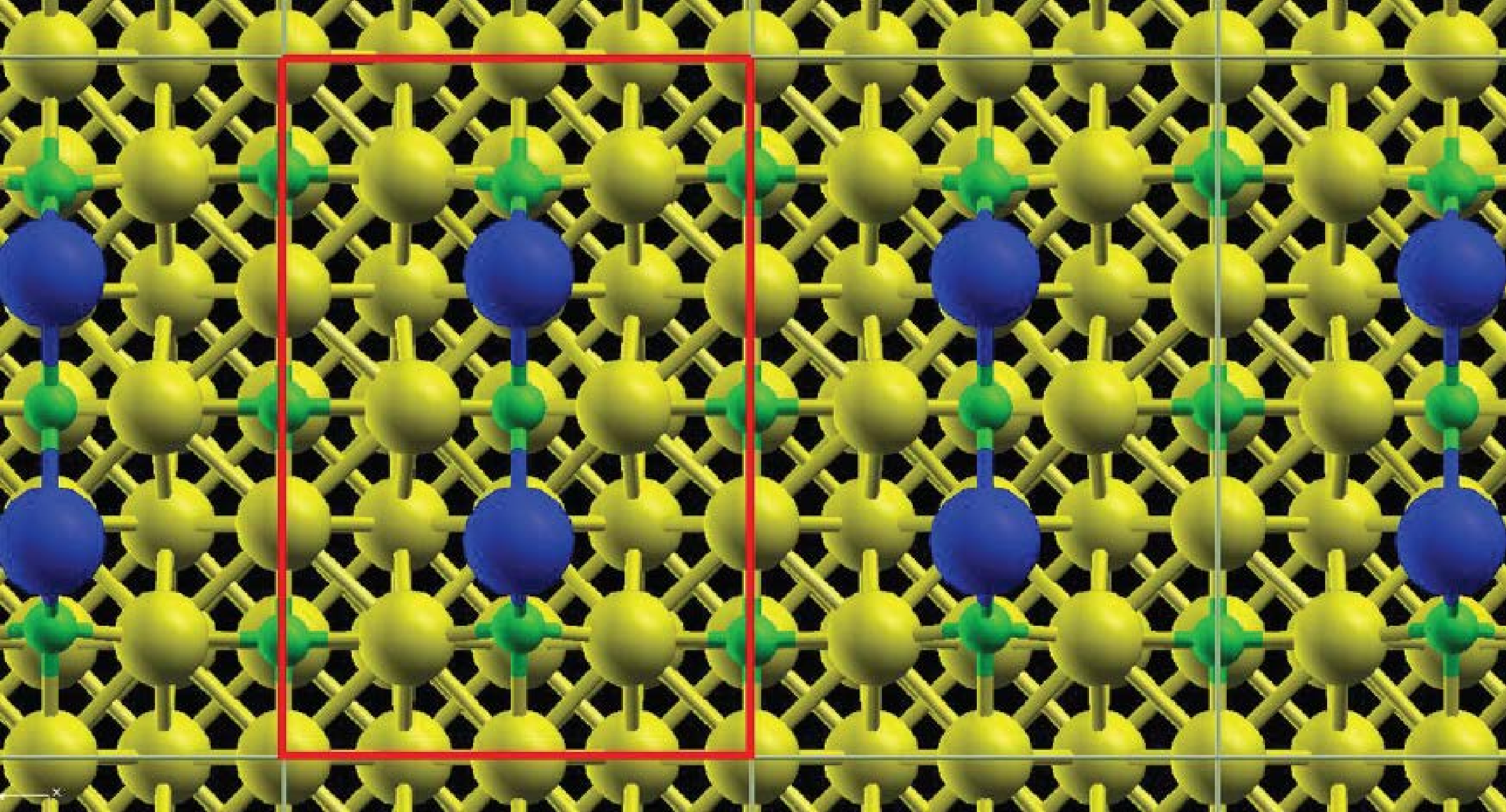}
\includegraphics[keepaspectratio,clip,height= 3.7cm,trim=0cm 0cm 0cm 0cm]{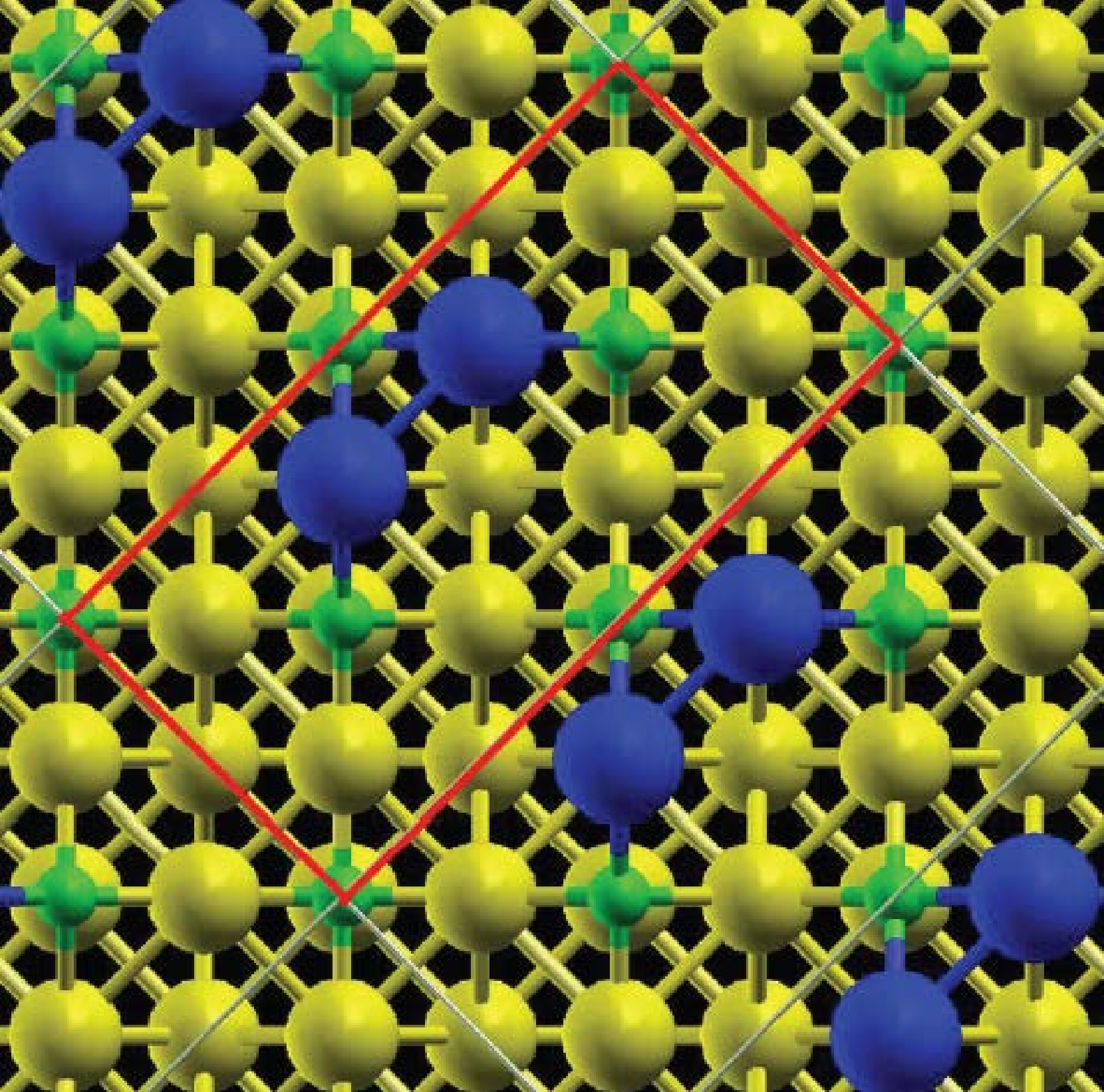}
\includegraphics[keepaspectratio,clip,height= 3.7cm,trim=0cm 0.2cm 0cm 0cm]{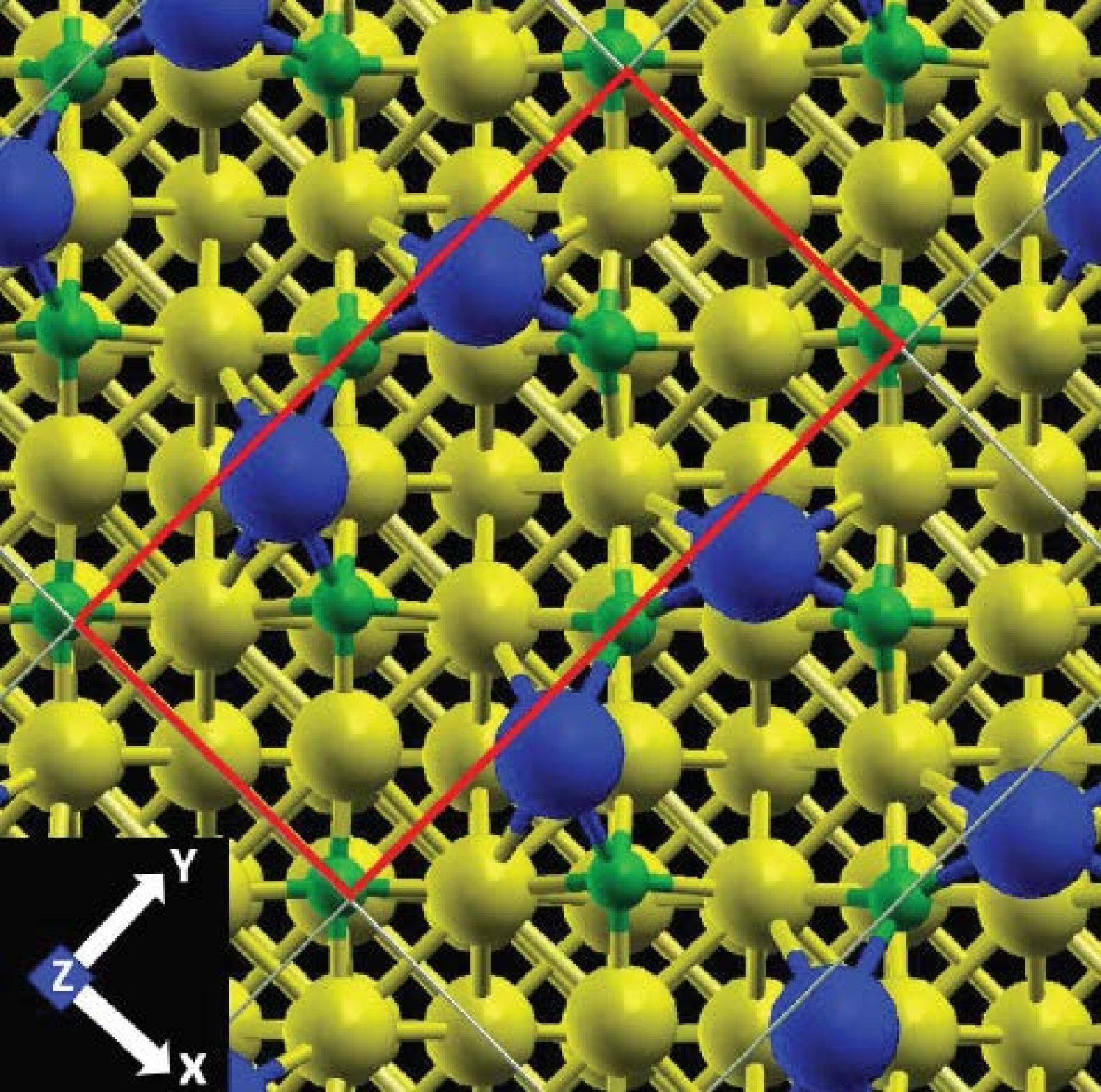}
\end{center}
\caption{\label{Gd2}
The upper figure shows the top view of the relaxed structure
of a linear Gd dimer on the CuN surface.
The lower left (right) figure shows the top view of the initial (relaxed) structure
of a diagonal Gd dimer on the CuN surface. Unit cells are marked by red rectangles.}
\end{figure}

In the STM experiments, a copper-nitride monolayer is built between a magnetic atom
and the Cu(100) surface \cite{MnScience,FeScience,FeCuScience,CoNaturePhysics,CoPRL}
to keep the atomic spin away from the screening of its underlying conduction electrons
while permitting a sufficient amount of STM tunneling current for probing the spin excitations.
To understand the magnetic properties of Gd atoms on the CuN surface,
we simulate a single Gd on this surface by first constructing
a supercell of 5-layer Cu slabs plus 8 vacuum layers with the nitrogen atoms snugging in-between
the half of vacant sites and then placing each Gd atom atop the Cu site
of the CuN surface within a $3\time3$ unitcell.
We perform density-functional calculations in the all-electron full-potential
linearized augmented plane wave (FLAPW) basis \cite{win2k}.
A naive local density approximation (LDA) or generalized gradient approximation (GGA),
when being applied to materials composed of rare-earth atoms,
generally yields $f$ levels inconsistent with photoemission experiments,
and needs to be fixed by adding extra on-site Coulomb repulsion
to the exchange-correlation functional, the so-called DFT+U method.
To determine the on-site Coulomb $U_f$ and exchange $J_f$ values of
the Gd $4f$ orbitals on the CuN surface,
we revisit the GdN bulk system, which mimics very well the local structure of Gd on the CuN surface.
The GdN bulk, being a ferromagnetic semiconductor,
has been studied experimentally by photoemission
and computationally by LSDA+U
for its potential application in spintronics.
Following the same way as the previous LSDA+U studies of GdN bulk
\cite{GdN-dftprl,GdN-dftapl},
we find that
$U_f=6.7$ eV and $J_f=0.7$ eV yield the energy difference between
the majority-spin Gd $4f$ and N $2p$ states
in best agreement with photoemission measurements \cite{GdN-expsurf,GdN-expprb}.
This set of $U_f$ and $J_f$ are used in our succeeding calculations of both
the GdN bulk and the Gd dimer on the CuN surface.

The spin couplings along the diagonal and the linear Gd-N-Gd paths of the GdN bulk
are well-studied in the literature \cite{GdN-dftapl,GdN-dftprb},
where the two Gd atoms along the paths are NN and next NN to each other, respectively,
as shown in Fig.~\ref{GdN}.
The two coupling paths between Gd atoms in a GdN bulk strongly
suggest that there are two possible geometries of Gd dimers on the CuN surface:
one has Gd-N-Gd along the N row, and the other in a right angle (two Gd atoms along the diagonal).
Previous GdN-bulk studies have concluded that the spin couplings between Gd atoms
of a GdN bulk along the diagonal and the linear paths are
ferromagnetic and antiferromagnetic, respectively.
Therefore we expect the surface Gd dimers in two geometries to have spin couplings
the same as their counterparts of similar geometries in the GdN bulk,
i.e. diagonal (linear) being ferromagnetic (antiferromagnetic).

\begin{figure}
\begin{center}
\includegraphics[keepaspectratio,clip,height= 6.2cm,trim=0cm 0cm 0cm 0cm]{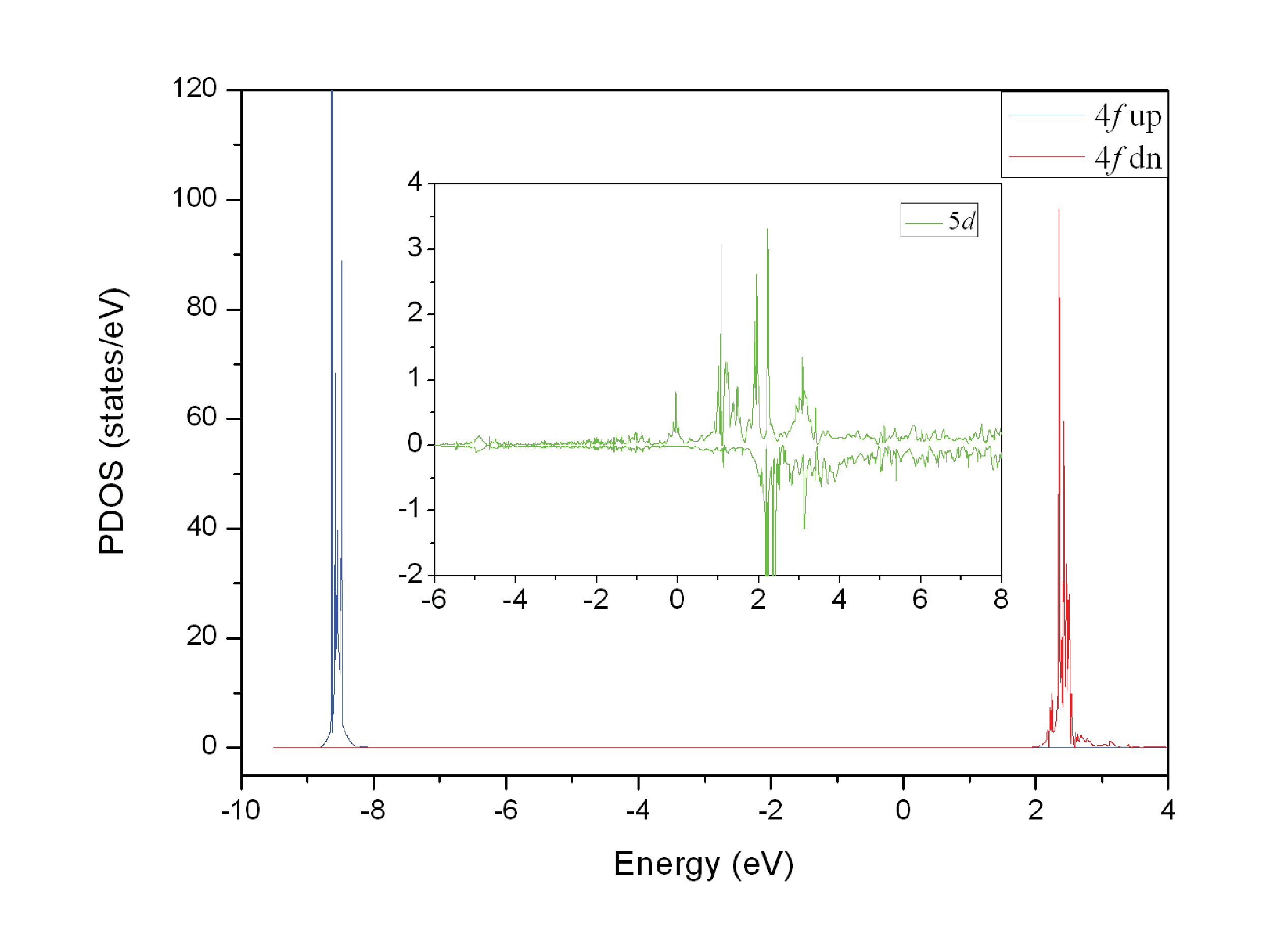}
\end{center}
\caption{\label{dos} Partial density of states of a single Gd on the CuN surface.}
\end{figure}

We therefore arrange Gd atoms in those two geometries on the CuN surface,
and optimize the crystal structures until the maximum force among all the atoms reduces to
$<\!\!\!\!\!\!\!\!_{_{_{\textstyle \sim}}}\,10$ mRy$/a_{0}$ and $5$ mRy$/a_{0}$.
The relaxed structures are shown in Fig.~\ref{Gd2},
and the dimer local geometries are quantitatively presented in Table \ref{Gd-N-Gd}.
It is interesting to notice that the diagonal Gd dimer relaxes its bond angle from $90^\circ$ to $112^\circ$.
This can be understood given that the diagonal Gd-to-Gd distance in GdN bulk is $3.52$\AA,
and the initial Gd-Gd distance on the surface is $2.56$\AA, much shorter than $3.52$\AA,
so a relaxed Gd-Gd distance on the surface of $3.64$\AA~is rather reasonable.
To determine the Gd spin on the CuN surface,
we plot the calculated partial density of states (PDOS)
of a single Gd on the CuN surface in Fig.~\ref{dos}.
One clearly sees that the $4f$ majority spin states are all occupied and the minority states
are all unoccupied, which implies a $4f^{7}$ configuration for Gd,
a spin-$7/2$ configuration for its $4f$ shell.
In addition, the $5d$ states are not occupied, and its rather small PDOS
in the entire energy range indicates its delocalization outside of the Gd atomic sphere,
in contrast to a free Gd atom that carries a valence configuration $5d^{1}4f^{7}6s^2$.
By comparing the Gd2/CuN and the GdN bulk,
we find that both systems have each Gd atom connecting to N atoms
such that their Gd local structures are very similar to each other
but are significantly different to that of a free atom.
The local structure plays an important role of the spins of all three Gd-contained systems.

To calculate the spin coupling $J$ between Gd spins on the CuN surface,
we take advantage of the correspondence between the collinear spins of a Heisenberg
model and the magnetic moments of the real crystal surface of interest \cite{MnDFT}.
The Hamiltonian of a Heisenberg spin dimer is
\begin{equation}
H=J {\bf S}_{1}\cdot{\bf S}_{2}.\label{Heisenberg}
\end{equation}
The difference of energy expectation values $\delta E$ between the parallel and antiparallel
spins is related to the coupling $J$, for spin-$S$ atoms, as
\begin{equation}
\delta E=JS^2-(-JS^2)=2JS^2 \label{EJS}
\end{equation}
By calculating the total energies of the parallel- and
antiparallel-spin configurations of
a Gd dimer at the Cu site of a CuN surface,
we obtain from (\ref{EJS}) the exchange coupling $J$ to
be $1.24$meV for the linear dimer and $-1.25$meV for the diagonal dimer.
Our calculations obtain an antiferromagnetic coupling for the linear Gd dimer
and ferromagnetic for the diagonal dimer.
One can also see in Table \ref{Gd-N-Gd} a trend of
the coupling $J$ varying with the angle $\angle$Gd-N-Gd.
The two Gd atoms start with a ferromagnetic coupling at an exact right angle,
becomes slightly less ferromagnetic at $\angle\mbox{Gd-N-Gd}=112^\circ$,
changes to antiferromagnetic at $\angle\mbox{Gd-N-Gd}=135^\circ$,
and finally stays antiferromagnetic when along a straight line.
\begin{table}
\begin{tabular}{|l|c|c|c|c|c|}
  \hline
                &  Gd to Gd (\AA)  &    Gd-N (\AA)   & $\angle$Gd-N-Gd  & $J$ (meV) \\
  \hline
   NN Gd    &  $3.52$ & $2.49$  &   $90^\circ$  & -1.51 \\
   in GdN bulk &&&& \\
  \hline
 next NN Gd &  $4.98$ & $2.49$  &   $180^\circ$ &  1.09 \\
   in GdN bulk &&&& \\
  \hline
 diagonal Gd dimer       &  $3.64$ & $2.19$  &   $112^\circ$ & -1.25 \\
   on CuN surface &&&& \\
  \hline
   linear Gd dimer       &  $4.15$ & $2.24$  &   $135^\circ$ & 1.24   \\
   on CuN surface &&&& \\
  \hline
\end{tabular}
\caption{\label{Gd-N-Gd} Calculated Gd-to-Gd distances, Gd-N bond length, Gd-N-Gd bond angle, and
the spin coupling $J$ between Gd of four systems.
}
\end{table}
\begin{table}
\begin{tabular}{|c|c|c|c|c|c|c|}
  \hline
   $ $    &   $J$  &  $J_\alpha$  & $J_\beta$  &  $J_d$  &  $J_s$  & $J_r$  \\
  \hline
  diagonal & $-1.25$ & $-0.92$ & $-1.30$ & $-0.97$ & $-0.33$ & $0.05$ \\
  \hline
  linear & $1.24$ & $-0.88$ & $0.66$ & $-1.46$ & $2.12$ & $0.58$\\
  \hline
\end{tabular}
\caption{\label{J} Calculated spin couplings in meV of
the original Gd dimers on the CuN/Cu(100) surface $J$,
the Gd dimers on a single CuN sheet $J_\beta$, and
$J_\alpha$ the spin coupling for the case of the N in between
the Gd substituted by a Ne atom.
Also listed are direct exchange $J_d$, superexchange $J_s$,
and RKKY $J_r$ extracted from $J$, $J_\alpha$, and $J_\beta$.}
\end{table}

\begin{figure}
\begin{center}
\includegraphics[keepaspectratio,clip,height= 3.1cm,trim=0cm 0cm 0cm 0cm]{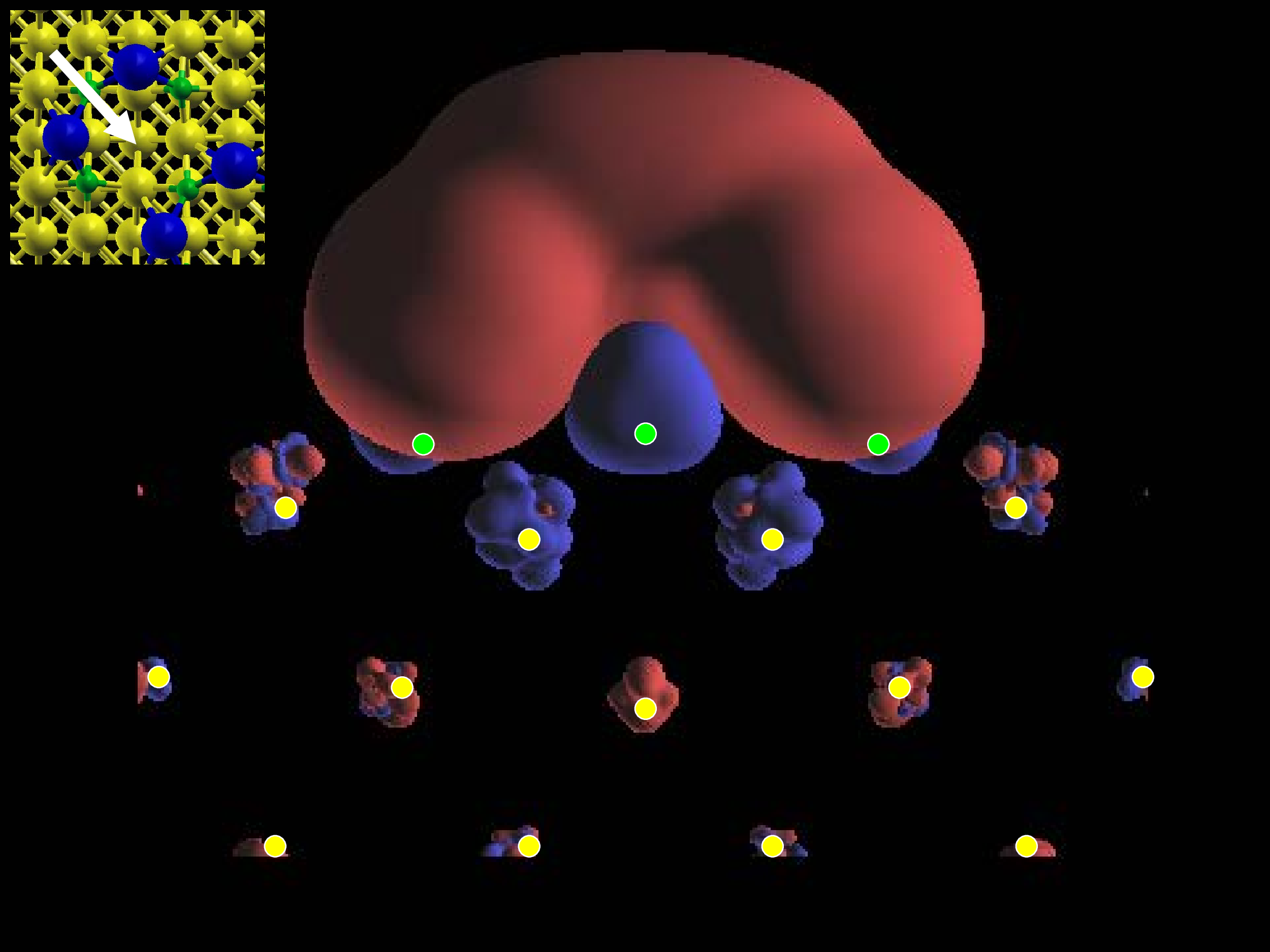}
\includegraphics[keepaspectratio,clip,height= 3.1cm,trim=0cm 0cm 0cm 0cm]{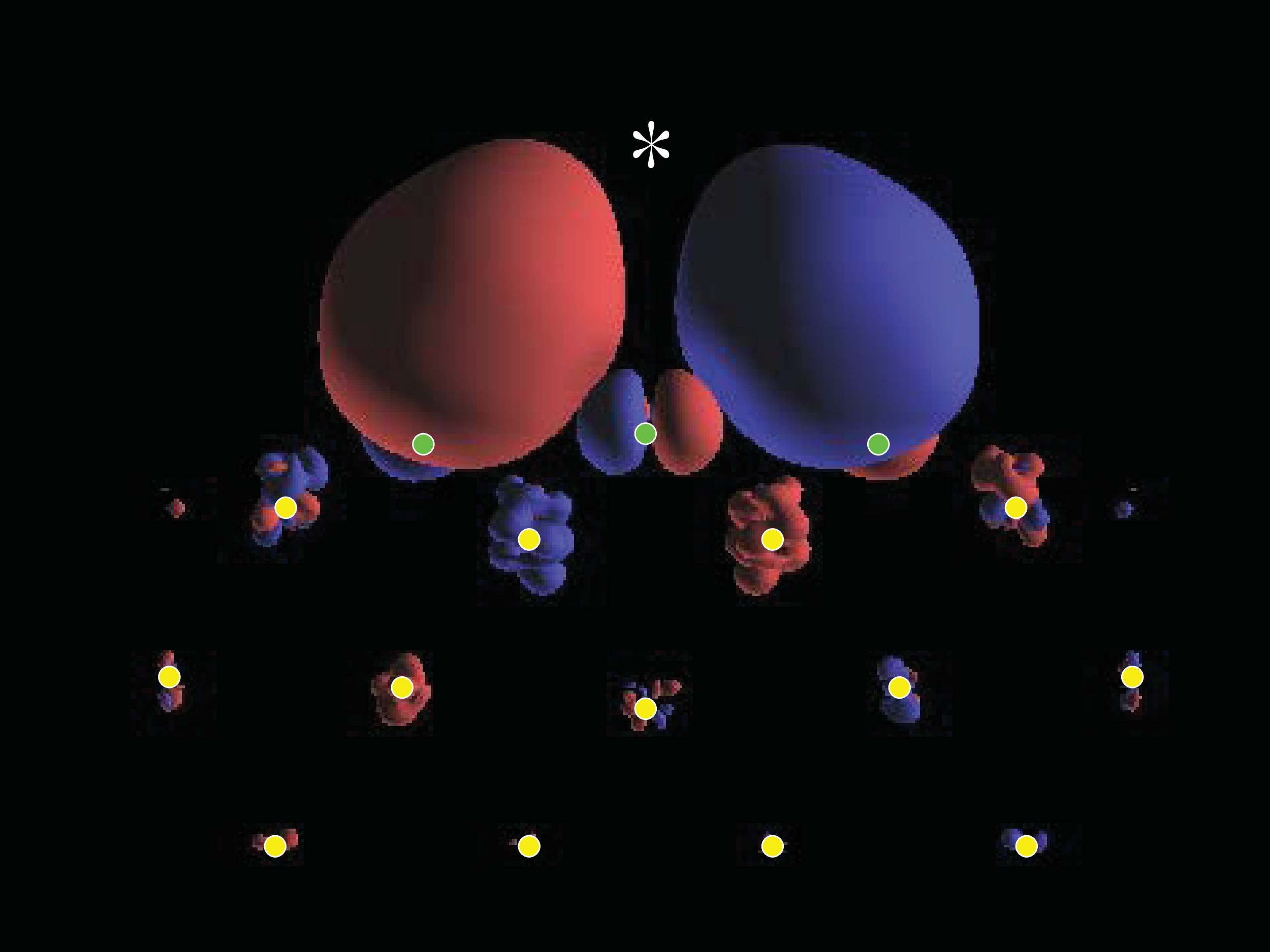}
\includegraphics[keepaspectratio,clip,height= 3.1cm,trim=0cm 0cm 0cm 0cm]{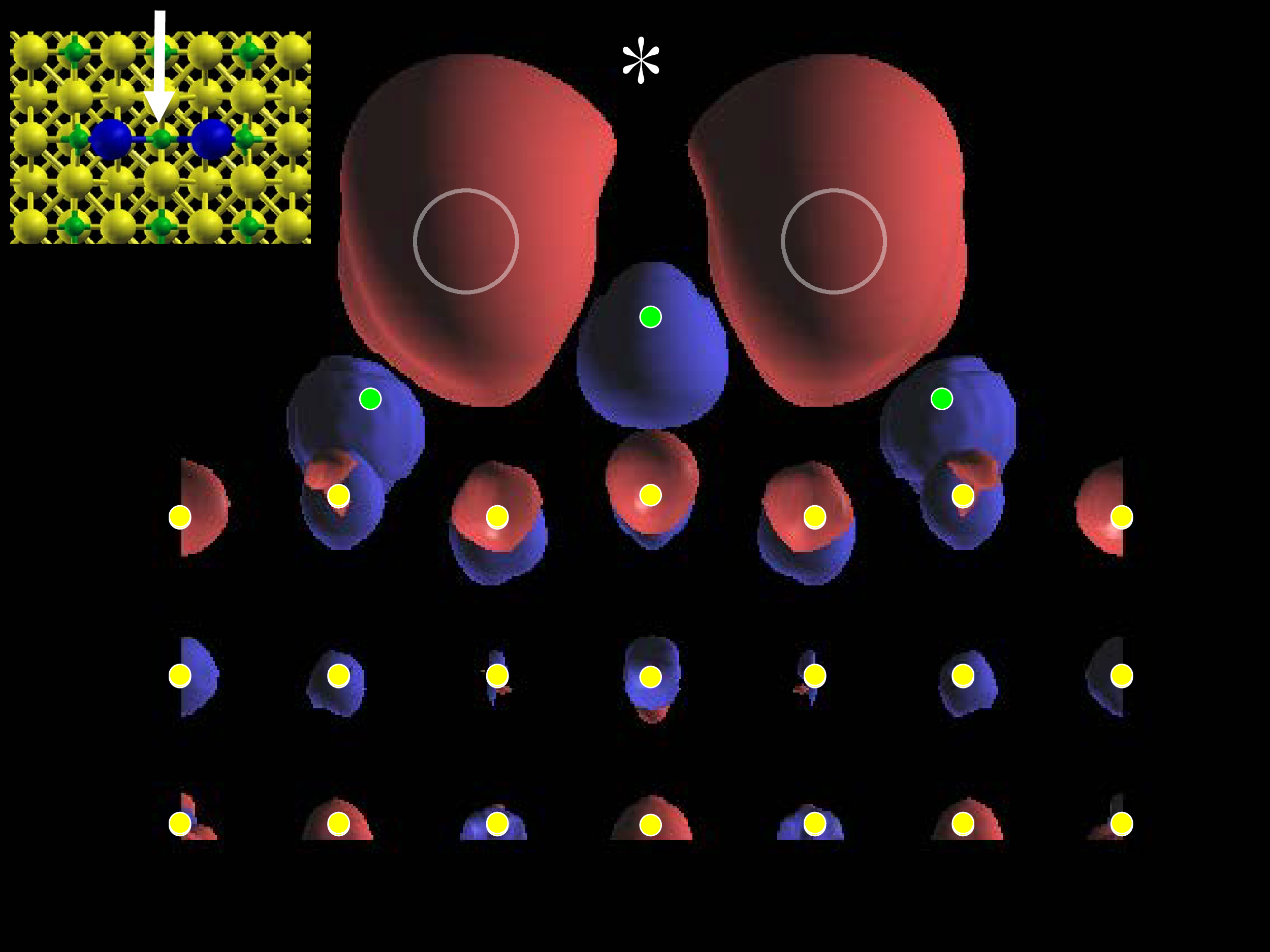}
\includegraphics[keepaspectratio,clip,height= 3.1cm,trim=0cm 0cm 0cm 0cm]{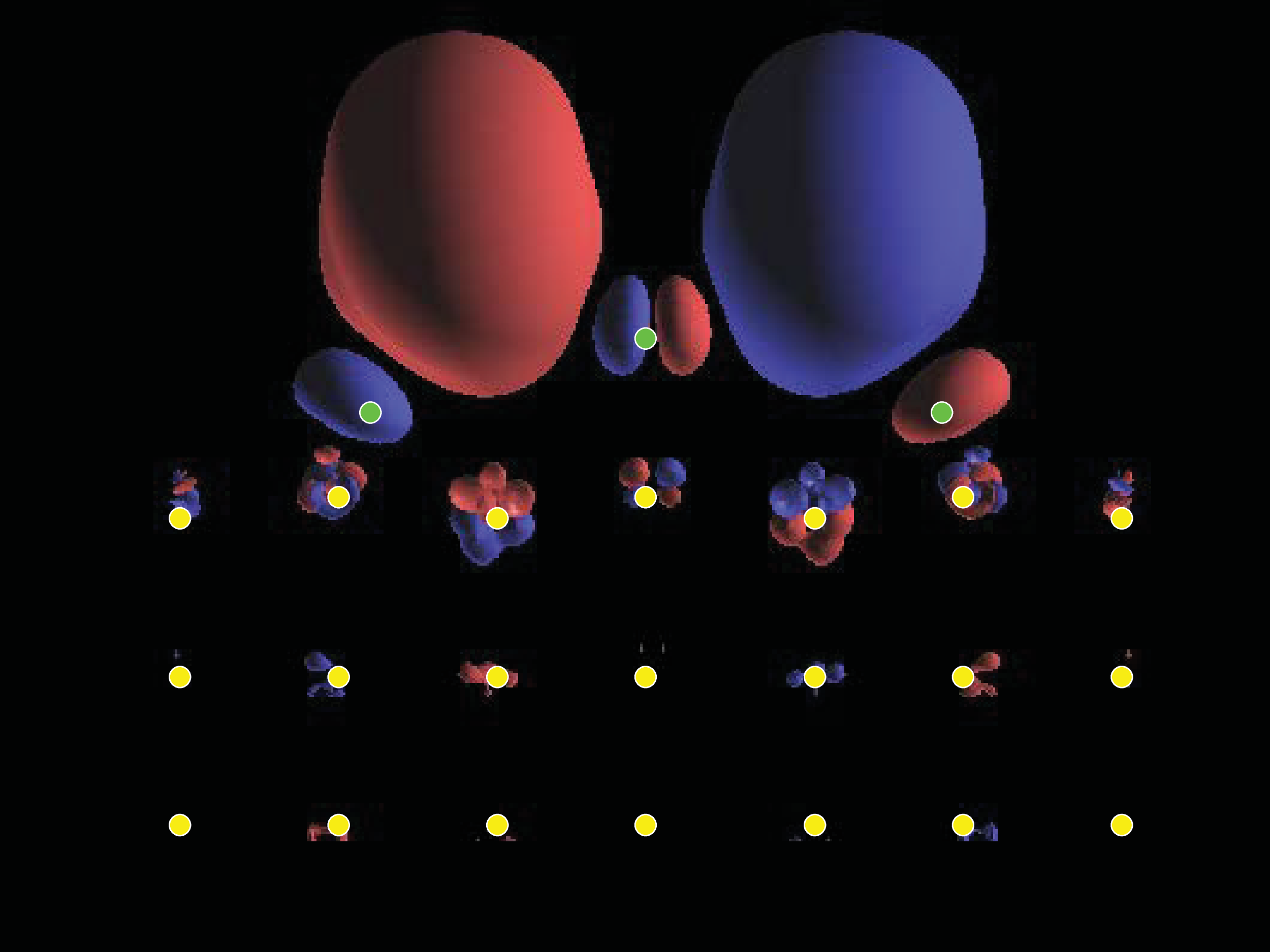}
\end{center}
\caption{\label{spin}
Calculated spin-density isosurfaces of diagonal (upper) and linear (lower) Gd dimers on CuN
in their ferromagnetic (left) and antiferromagnetic (right) configurations.
Red stands for positive magnetization, and blue for negative.
The white arrow in the stick-ball inset of each plot indicates the observation direction.
The green and yellow solid circles denote the positions of N and Cu, respectively.
All spin densities are plotted at the magnitude of $0.001e/$\AA$^3$.
A ``$\ast$'' symbol is labeled in the less stable spin configuration of each Gd dimer geometry.}
\end{figure}

We now turn our attention to the underlying mechanism
of the ferromagnetism (antiferromagnetism) of the diagonal (linear) Gd dimer.
There are three possible magnetic interactions that
may couple the two Gd spins: direct exchange,
superexchange, and the Ruderman-Kittel-Kasuya-Yosida (RKKY) interaction.
To extract the three components out of the resultant coupling,
we perform calculations of two alternative model systems.
One is the original Gd dimer on the surfce with the in-between N atom
replaced by a Ne atom,
which effectively turns off the superexchange.
The other is a Gd dimer on top of a single CuN monolayer, i.e.,
removing the underlying metallic Cu(100) slab,
with the dimer and CuN sheet remaining, which has basically no RKKY.
The atomic positions of these two alternative model systems
exactly follow the original realistic surface, i.e. their structures are not relaxed,
so that the magnetic coupling is the only difference among these systems.
We then decompose the spin couplings of the three systems:
the original Gd dimer on the CuN/Cu(100) surface $J$,
the one with a Ne in between two Gd atoms $J_\alpha$,
and the Gd dimer on a single CuN sheet $J_\beta$, into the contributions
of direct exchange $J_d$, superexchange $J_s$, and RKKY $J_r$.
We write down their relations as
\begin{eqnarray}
&&J=J_d+J_s+J_r, \nonumber \\
&&J_\alpha=J_d+J_r, \nonumber \\
&&J_\beta=J_d+J_s.
\end{eqnarray}
It is a simple matter to solve for $J_d$, $J_s$, and $J_r$.
The calculated spin couplings are listed in Table \ref{J}.

For the diagonal dimer, the insignificant differences
among $J$, $J_\alpha$, and $J_\beta$
imply that the N atom in between and the underlying conduction
electrons play minor roles in the spin coupling of the Gd dimers,
while the direct wavefunction overlapping between the two Gd atoms
actually dominates. In fact, the obtained $J_d$, $J_s$, and $J_r$ values
of the diagonal dimer reflect the statement above,
where the superexchange and RKKY are $34\%$ and $5\%$
of the direct exchange, respectively.
Notably, we find a ferromagnetic superexchange.
The linear dimer, on the other hand,
has quite different components of the three types of magnetic couplings.
Antiferromagnetic superexchange dominates, larger than the direct
exchange, and there is also significant antiferromagnetic RKKY,
at $27\%$ of superexchange.
When comparing the same types of magnetic interaction between the two dimer geometries,
we find that the direct exchange of both dimers have the same sign and order of magnitude
while their superexchange $J_s$ have opposite signs.
Consequently, the magnitude of $J_s$ determines the sign of the dimers' total $J$.
With the knowledge that the ferromagnetic coupling for the diagonal case is partly due
to ferromagnetic superexchange, we revisit Table \ref{Gd-N-Gd}, and interpret the
trend of the coupling $J$ varying with the angle $\angle$Gd-N-Gd, as a rough measure of the
angle dependence of superexchange in these systems.
The prediction of ferromagnetic superexchange at 90 degrees,
varying through a sign change and ending up at antiferromagnetic
for 180 degrees is well known in the chemistry community \cite{fmsuperex},
and is due to different orbitals being involved
with the superexchange hopping to the N at different angles.
Pauli exclusion favors antiferromagnetic coupling
only when the three atoms are more or less in a line.
At sharper angles, the relative symmetries of the orbitals
involved in the hopping through the N tend to favor ferromagnetic alignment of the spins.

\begin{figure}
\begin{center}
\includegraphics[keepaspectratio,clip,height= 3.1cm,trim=0cm 0cm 0cm 0cm]{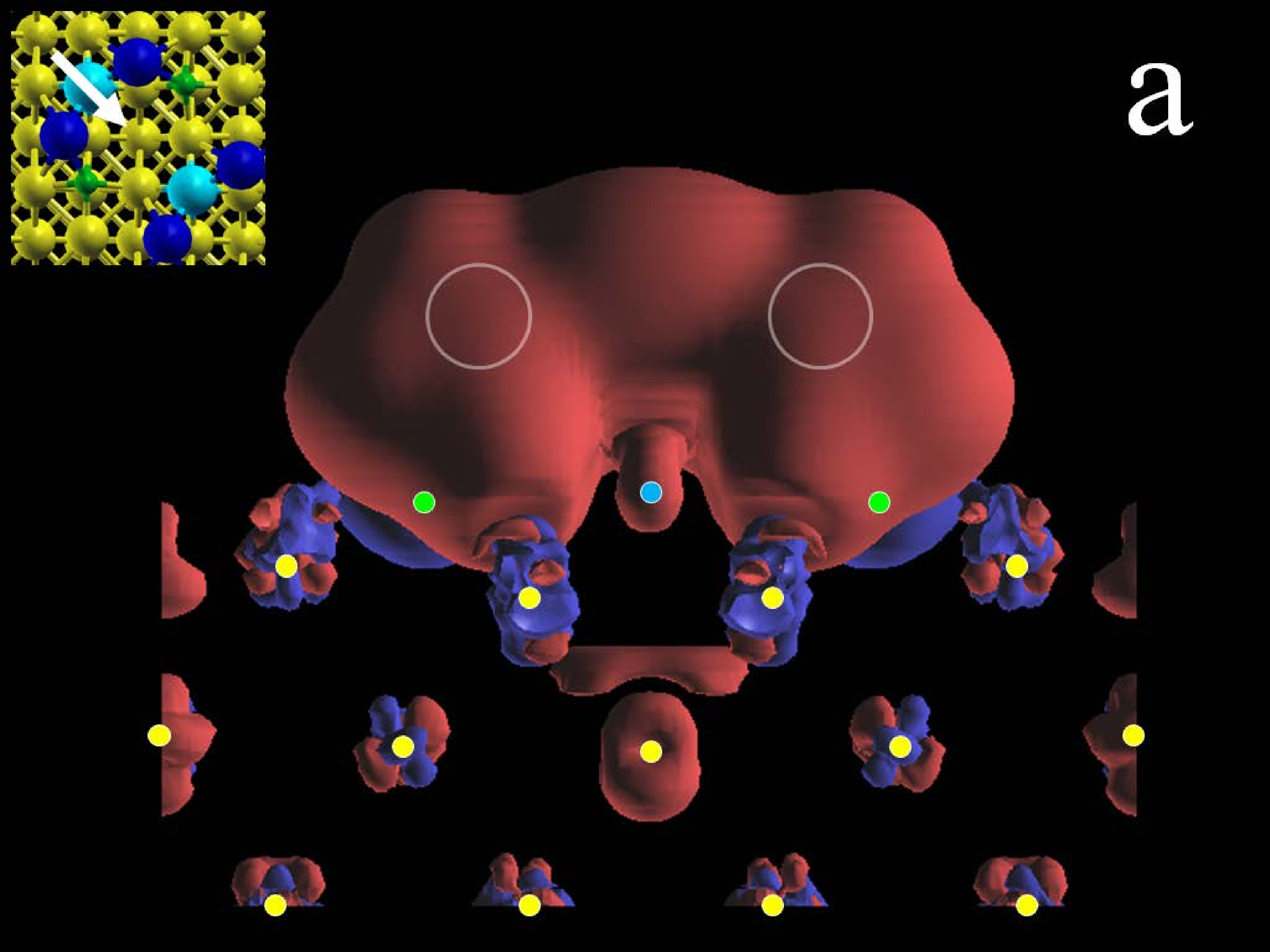}
\includegraphics[keepaspectratio,clip,height= 3.1cm,trim=0cm 0cm 0cm 0cm]{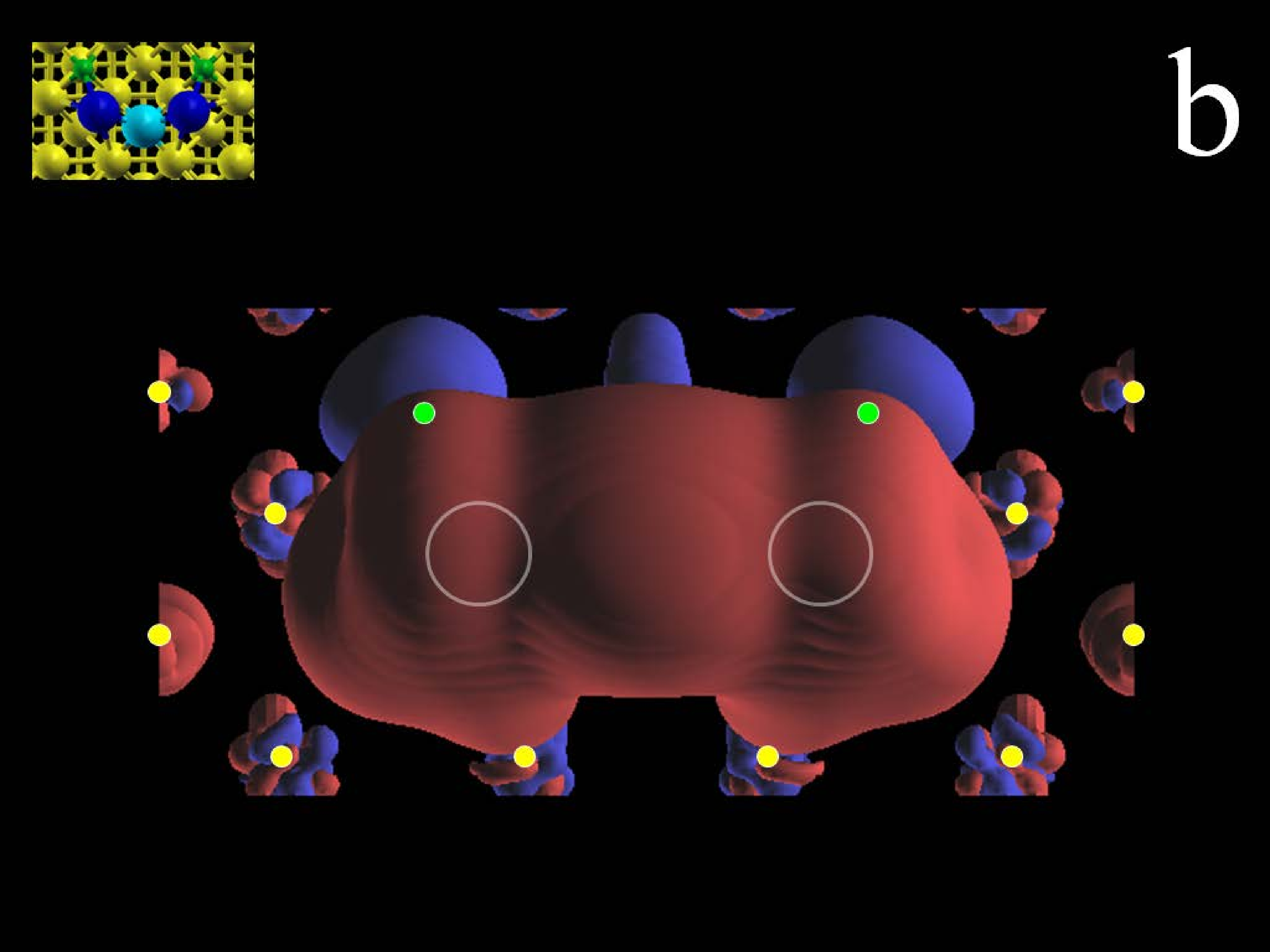}
\includegraphics[keepaspectratio,clip,height= 3.1cm,trim=0cm 0cm 0cm 0cm]{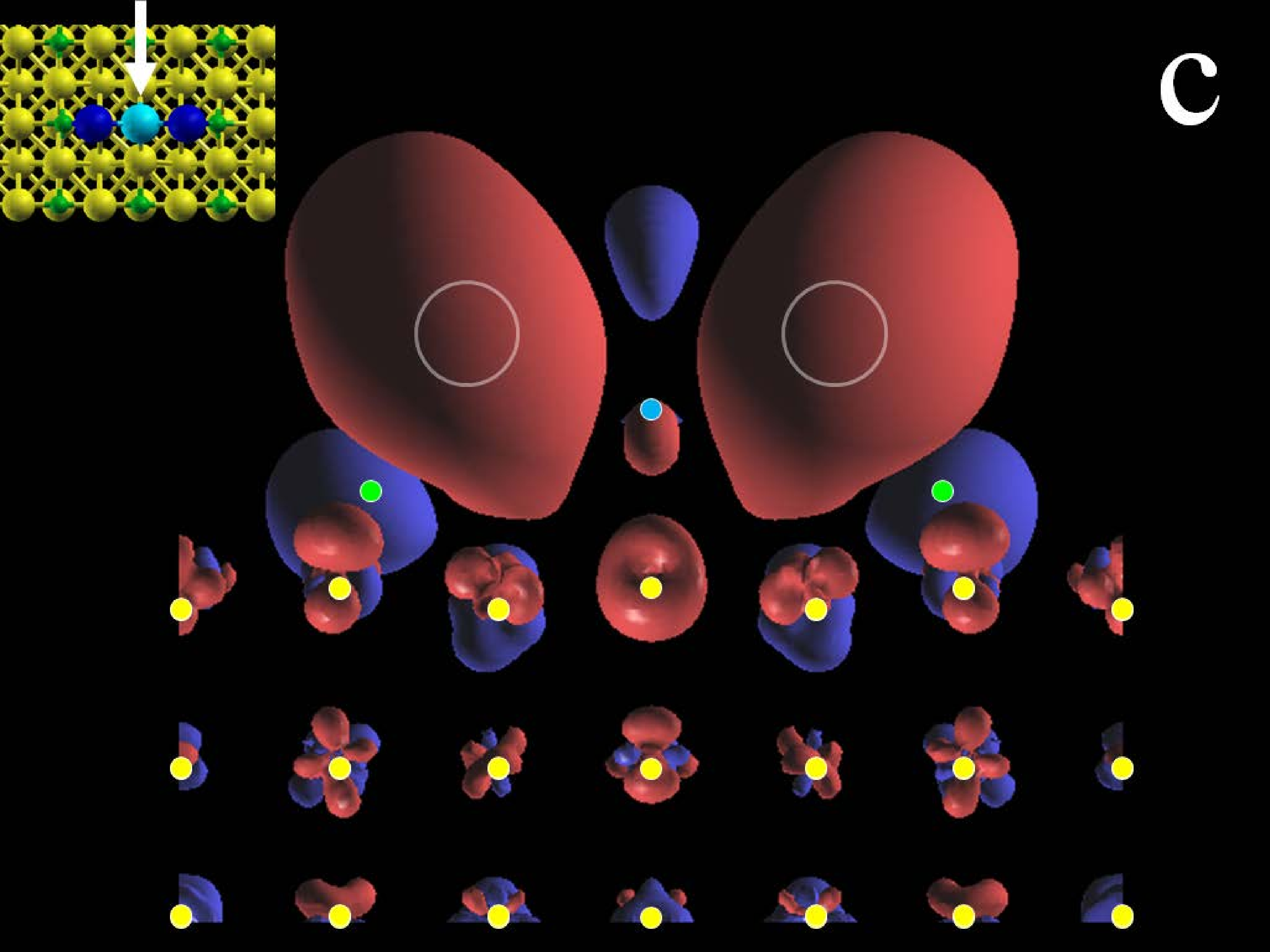}
\includegraphics[keepaspectratio,clip,height= 3.1cm,trim=0cm 0cm 0cm 0cm]{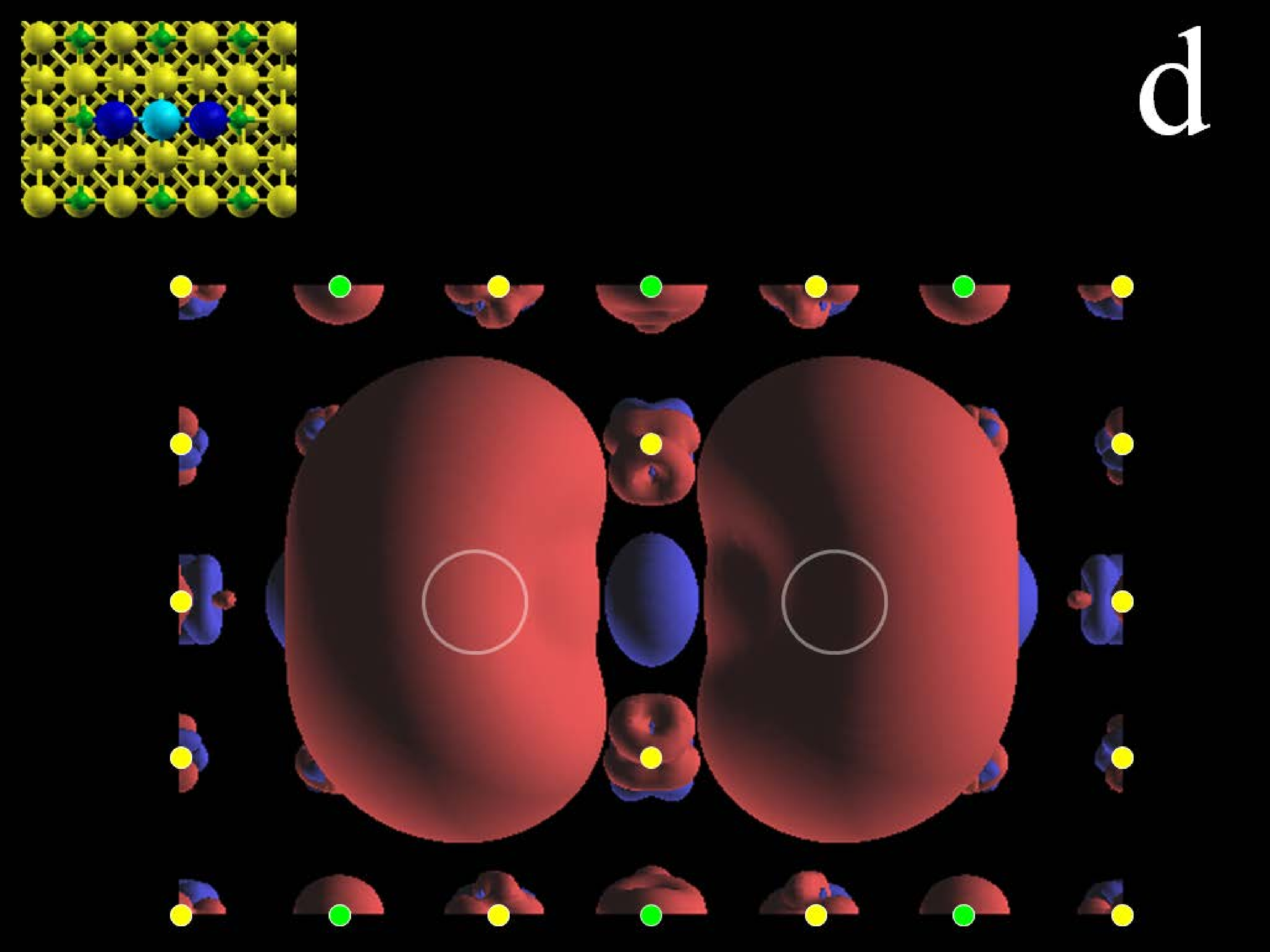}
\end{center}
\caption{\label{Nespin}
Calculated spin-density isosurfaces of the modeled Gd dimers
on CuN with the in-between N replaced by Ne.
Such dimers will be called Gd-Ne-Gd dimer later,
and both are in their ferromagnetic configuration here.
({\bf a}) and ({\bf b}) The side and top views of the diagonal.
({\bf c}) and ({\bf d}) The side and top views of the linear.
For all four plots, red stands for positive magnetization, and blue for negative.
The white arrow in the stick-ball inset of each side-view plot indicates the observation direction.
The green, yellow, and light blue solid circles denote the positions of N, Cu, and Ne, respectively.
All spin densities are plotted at the magnitude of $0.0005e/$\AA$^3$.}
\end{figure}

In order to obtain more physical insights for the Gd dimers,
we plot their spin densities for both the parallel- and antiparallel-spin
configurations in Fig.~\ref{spin}.
The parallel-spin configuration of a diagonal dimer has its spin density
forming one isosurface lobe, while the antiparallel-spin
has two disconnected Gd lobes of opposite spins.
In the ferromagnetic configuration,
the intermediate N spin is partially enveloped by the Gd-dimer spin lobe,
and is carrying an oppositely polarized spin.
In the antiferromagnetic configuration, the N atom in between becomes
a magnetic dipole antisymmetrically polarized by the two opposite Gd spins.
We compare this spin density with that of a linear Gd dimer
on the same surface in Fig.~\ref{spin}.
We see that while the ferromagnetic diagonal dimer has
its two spin lobes connected with each other,
the ferromagnetic linear dimer forms two disjoint ones.
The linear dimer has an antiferromagnetic ground state,
and the corresponding spin density has a nodal plane exactly in the middle of the two Gd.
Direct spin interchange in the linear case is expected to be less strong
because of the intervening Nitrogen, and compared to
the ``spin bonding'' between the two Gd atoms in the diagonal case,
the latter implying a strong overlapping of their spin unpaired orbitals.
However, from Table \ref{J} we see that the linear case
has direct exchange $1.5$ times that of the diagonal.

\begin{figure}
\begin{center}
\includegraphics[keepaspectratio,clip,height= 6.5cm,trim=0cm 0cm 0cm 0cm]{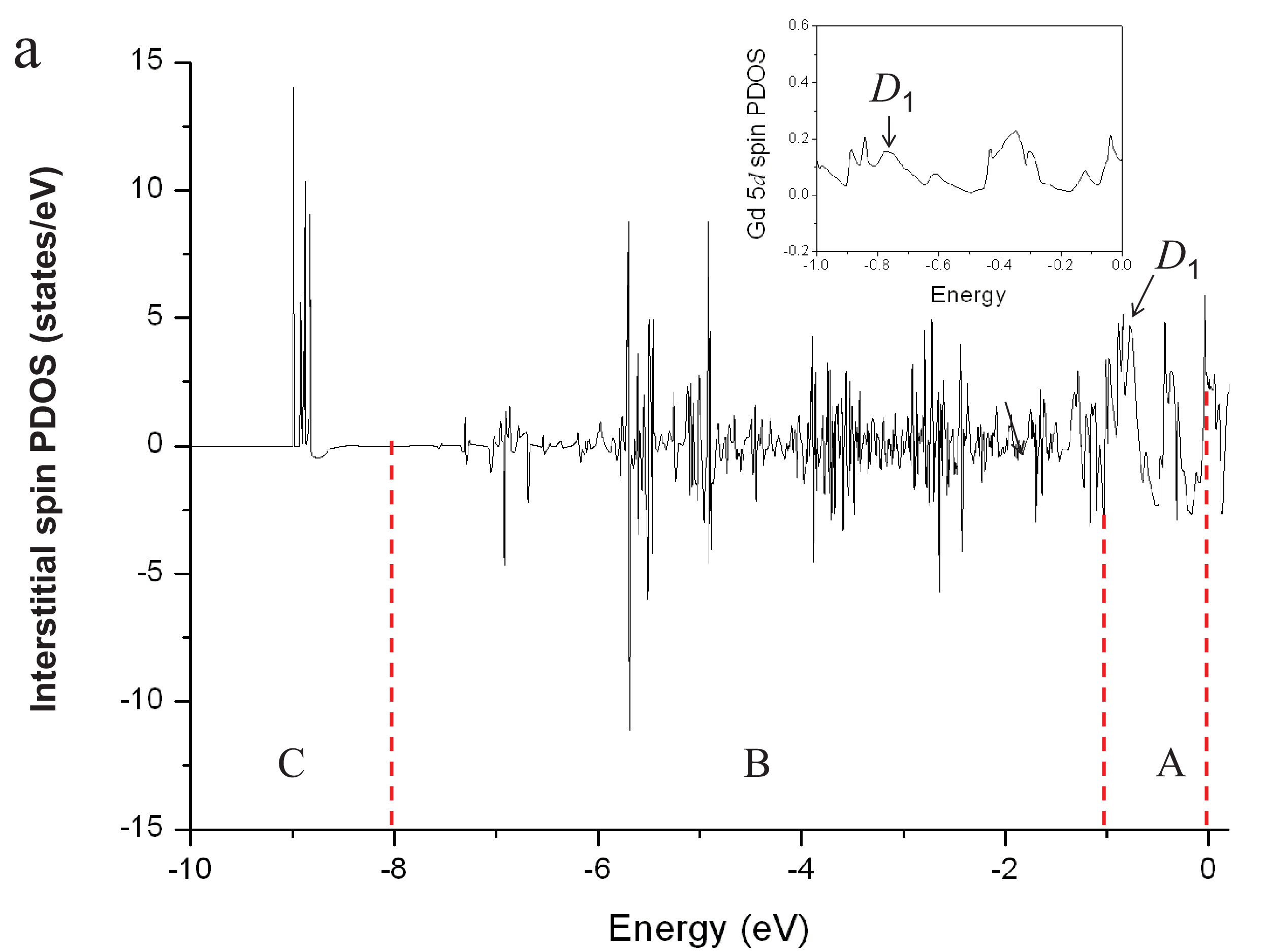}
\includegraphics[keepaspectratio,clip,height= 6.5cm,trim=0cm 0cm 0cm 0cm]{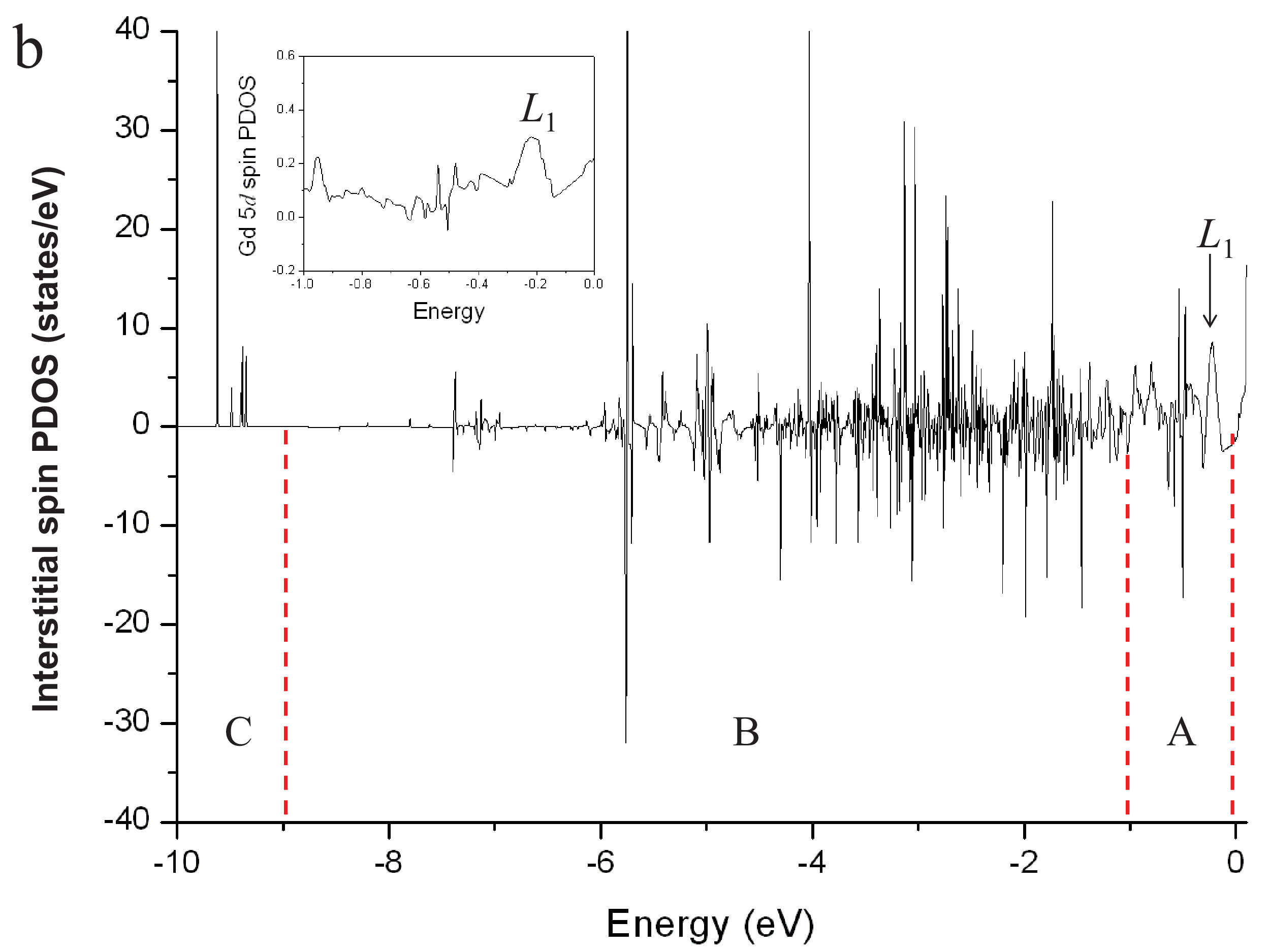}
\includegraphics[keepaspectratio,clip,height= 3.1cm,trim=0cm 0cm 0cm 0cm]{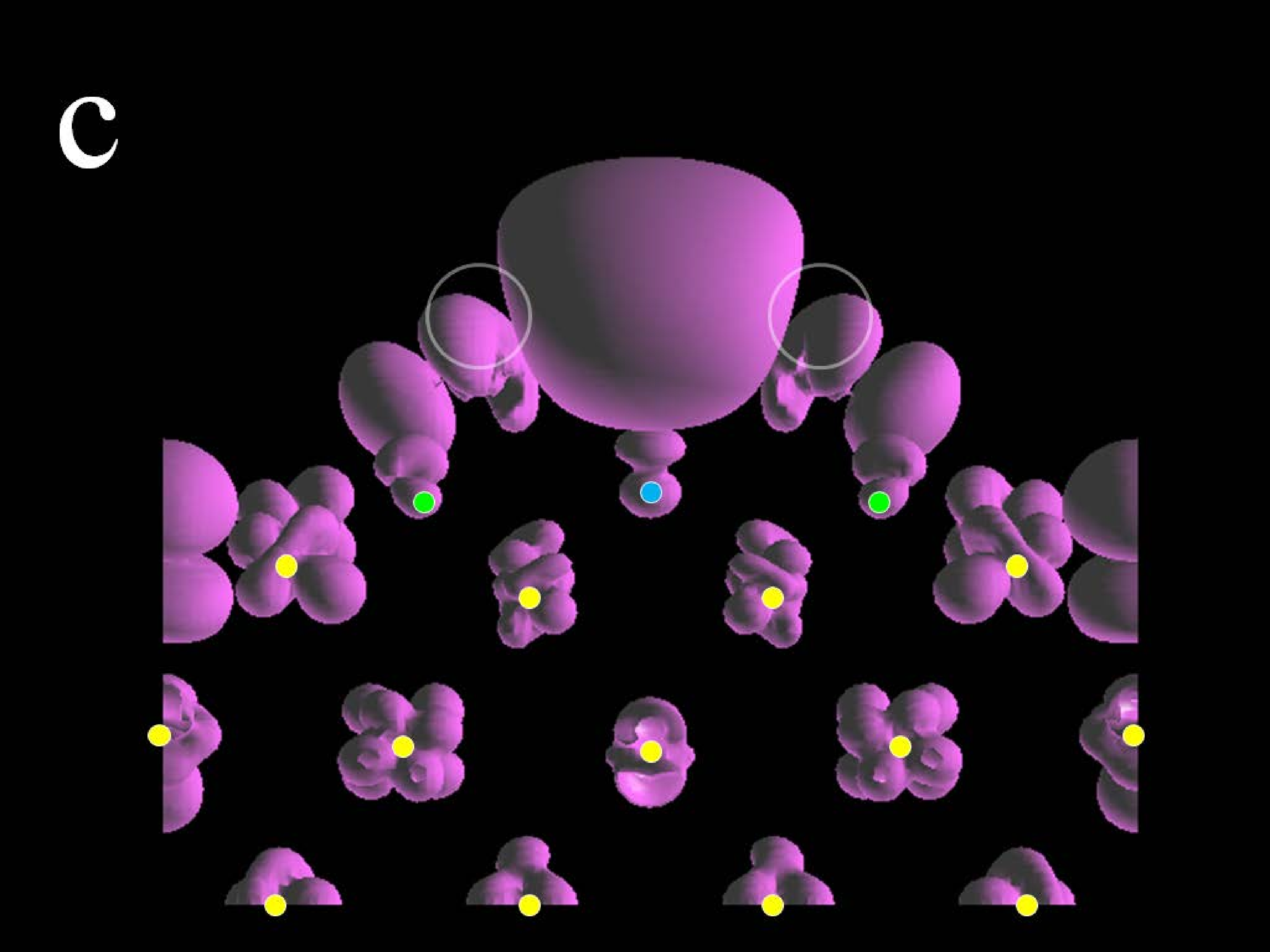}
\includegraphics[keepaspectratio,clip,height= 3.1cm,trim=0cm 0cm 0cm 0cm]{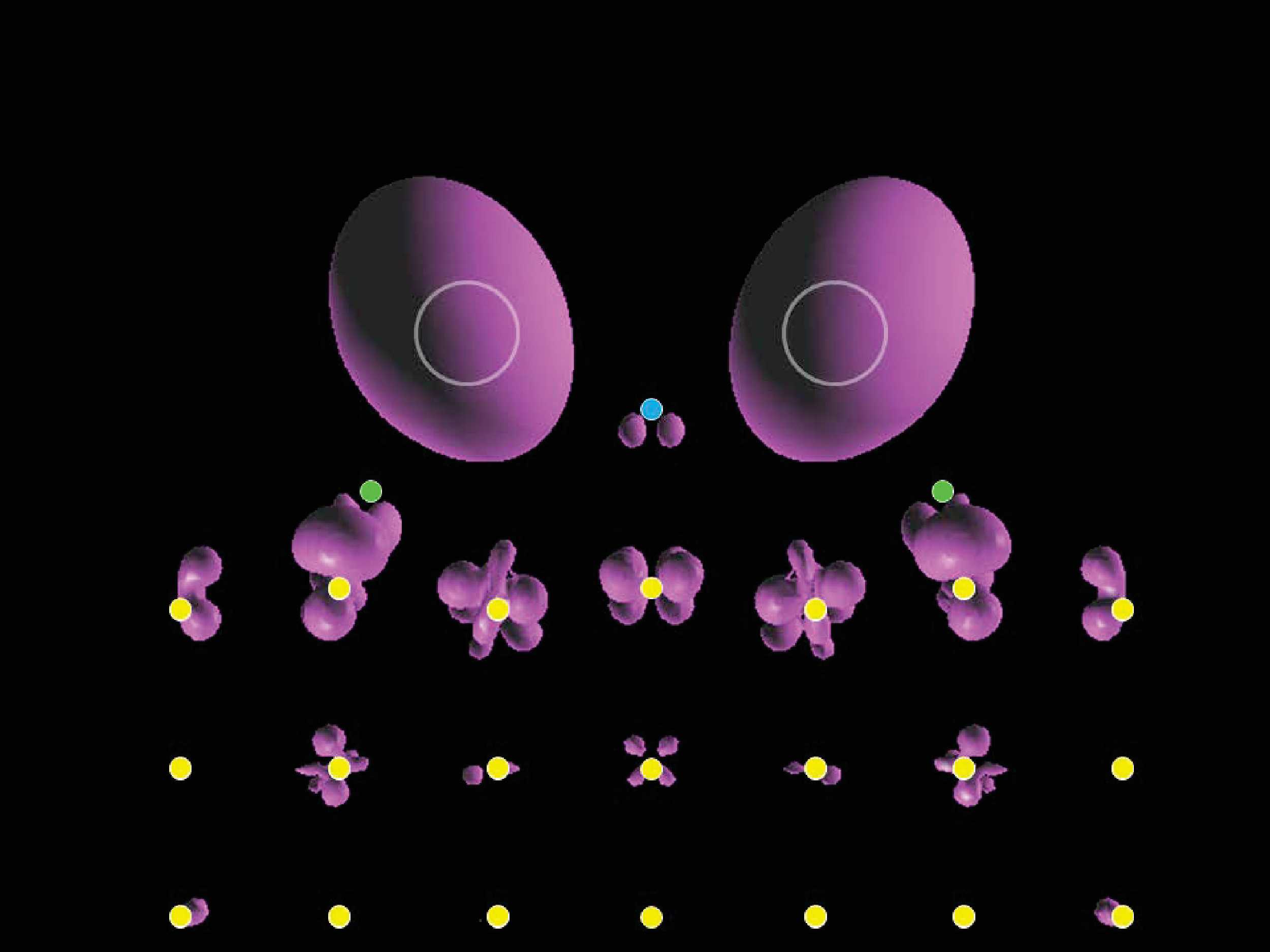}
\end{center}
\caption{\label{Ne_diagset}
({\bf a}) Spin-dependent partial density of states (S-PDOS)
in the entire interstitial region of a ferromagnetic diagonal Gd-Ne-Gd dimer on CuN,
defined as the majority-spin DOS minus the minority-spin in this spatial region.
The plotted energy range is divided into three regions.
Region A: the S-PDOS is mostly positive.
The peak that dominates the S-PDOS is labelled as ``$D_1$''.
Region B: the S-PDOS is highly oscillating.
Region C: a single positive peak at the energy of the majority-spin $4f$ level.
The inset shows the Gd $5d$ S-PDOS in the energy range A.
({\bf b}) Same plots as ({\bf a}) for the linear Gd-Ne-Gd dimer.
The dominating peak is labelled as ``$L_1$''.
({\bf c}) The majority-spin probability densities of the orbitals $D_1$ (left) and $L_1$ (right),
viewed along the same orientation as the side view of the spin density of
the diagonal and linear Gd-Ne-Gd dimers in Fig.~\ref{Nespin}, respectively.
The isosurfaces are plotted at the value of $0.02$\AA$^{-3}$.}
\end{figure}

In order to understand the physics underlying the direct exchange in these two configurations,
we set out to isolate the direct exchange.
The strong superexchange of the linear dimer results in an antiferromagnetic ground state,
and the dimer has a fundamental nodal plane in the middle, which
obscures the underlying direct exchange in the spin density.
To eliminate the superexchange from the system, we study instead
our model systems with N replaced by Ne.
Both superexchange-free configurations exhibit a ferromagnetic ground state,
which we attribute to the underlying direct exchange.
We plot their ferromagnetic spin densities in Fig.~\ref{Nespin}.
The diagonal Gd-Ne-Gd dimer has a spin-density lobe very
similar to that of a realistic Gd dimer,
which is consistent with the dominant direct exchange seen in Table \ref{J}.
The linear Gd-Ne-Gd dimer, on the other hand, provides new information from its shape.
We see a negative spin lobe exactly in the middle, together with
prolongation of positive polarization perpendicular to the dimer, as viewed from the top.
In both configurations we see the strong effect of orbital symmetry in the interactions.
In the diagonal case there are no nodes, and the linear case we see there are two nodes,
with a small maximum in the middle. This strong indication of orbital interaction is
further confirmation that we are seeing direct exchange.

In order to understand further the details of the interstitial spin density between
the Gd atoms of the diagonal dimer,
we plot the spin-dependent partial density of states (S-PDOS)
in the entire interstitial region of the unit cell (Fig.~\ref{Ne_diagset}a)
where S-PDOS is defined as the minority-spin DOS subtracted
from the majority-spin in this region.
There is an isolated positive peak at $E=-9$ eV,
which is exactly the energy of the majority-spin $4f$ level.
The highly oscillating S-PDOS within the energy range from $E=-8$ to $-1$ eV
results in the spins of the majority- and
minority-spin orbitals approximately cancelling.
In contrast, from $E=-1$ eV to the Fermi energy,
the S-PDOS shows a peak that dominates, labelled ``$D_1$''.
$D_1$ actually represents three equivalent near-energy peaks.
There are also two other peaks in region A.
In Fig.~\ref{Ne_diagset}c,
we plot the majority-spin probability density of $D_1$.
We can see that the highest probability of $D_1$ is mainly concentrated in
the interstitial region in between two Gd, and
provides the major contribution to the interstitial spin density.
Although we do not plot the other two near-energy peaks here,
we find that they look very similar.
In contrast, all other peaks, positive and negative,
in region A have negligible weight between the two Gd,
and represent the interstitial electrons in the bulk Cu.
The interstitial contribution of $D_1$ is decomposed into plane waves in the FLAPW basis.
To trace the atomic configuration of the above interstitial spin between the Gd,
we calculate the projections of the $D_1$ orbital to
$s$, $p$ and $d$ symmetries within a Gd muffin-tin sphere.
(There is no appreciable $f$ DOS in the interstitial region.)
The calculated muffin-tin projections to $6s$, $6p$, and $5d$ are
$17\%$, $13.5\%$, and $68.5\%$ of Gd, respectively.
We also plot the Gd $5d$ S-PDOS around the $D_1$ energy,
and find that there is also a peak exactly at $D_1$.
Consequently we believe that the Gd $5d$ electrons are delocalized across the diagonal
to enable the direct exchange.

We also perform S-PDOS analysis for the linear Gd-Ne-Gd dimer,
and plot it in Fig.~\ref{Ne_diagset}b.
The linear-dimer S-PDOS can be analyzed in a manner similar to that of the diagonal.
There are a number of peaks in region A.
However, the peak that dominates the S-PDOS surrounding the Gd
is the one which we label ``$L_1$'' in the figure.
Unlike the diagonal Gd-Ne-Gd dimer,
the $L_1$ peak has a wide width and does not have energy neighbors.
The majority-spin probability is also different from $D_1$,
and is elongated on the surface perpendicularly to the Gd-Gd direction,
very similarly to Fig.~\ref{Nespin}d.
As seen in Fig.~\ref{Ne_diagset}c,
we notice that the two Gd orbital lobes of $L_1$, unlike the $D_1$,
do not connect with each other, and instead have a node in between
and an antibonding appearance.

As calculated for $D_1$, the muffin-tin projections of $L_1$ to Gd $6s$, $6p$, and $5d$
are $33\%$, $3\%$, and $64\%$ of Gd, respectively.
The $5d$ character of $L_1$ within the Gd muffin-tin sphere is consistent with
the $5d$ S-PDOS plotted in the inset of Fig.~\ref{Ne_diagset}b.
Therefore, we also believe that the Gd $5d$ electrons, as in the diagonal case,
are delocalized to enable the direct exchange of the linear dimer.
The spreading of $L_1$ out of the Gd muffin-tin spheres into the interstitial region
represents $54\%$ of the total probability distribution, while $D_1$ plus its two near-energy
neighbors represent only $43\%$.
We see the linear configuration has a contribution roughly $1.3$ times that of the diagonal,
a ratio close to the $1.5$ ratio between the direct exchange couplings
of the two configurations shown in Table \ref{J}.
In fact, if we quantify the interstitial magnetic moments of
the diagonal and linear Gd-Ne-Gd dimers, they are $1.53$ and $2.34$, respectively.
with a ratio 1.54, even closer to $1.5$.
The absence and presence of a middle nodal plane in between the two geometries of the Gd-Ne-Gd dimers
are very likely due to the symmetry of their (hybridized) delocalized Gd orbitals,
a coupling we refer to generically as spin bonding.

\begin{figure}
\begin{center}
\includegraphics[keepaspectratio,clip,height= 4.1cm,trim=0cm 0cm 0cm 0cm]{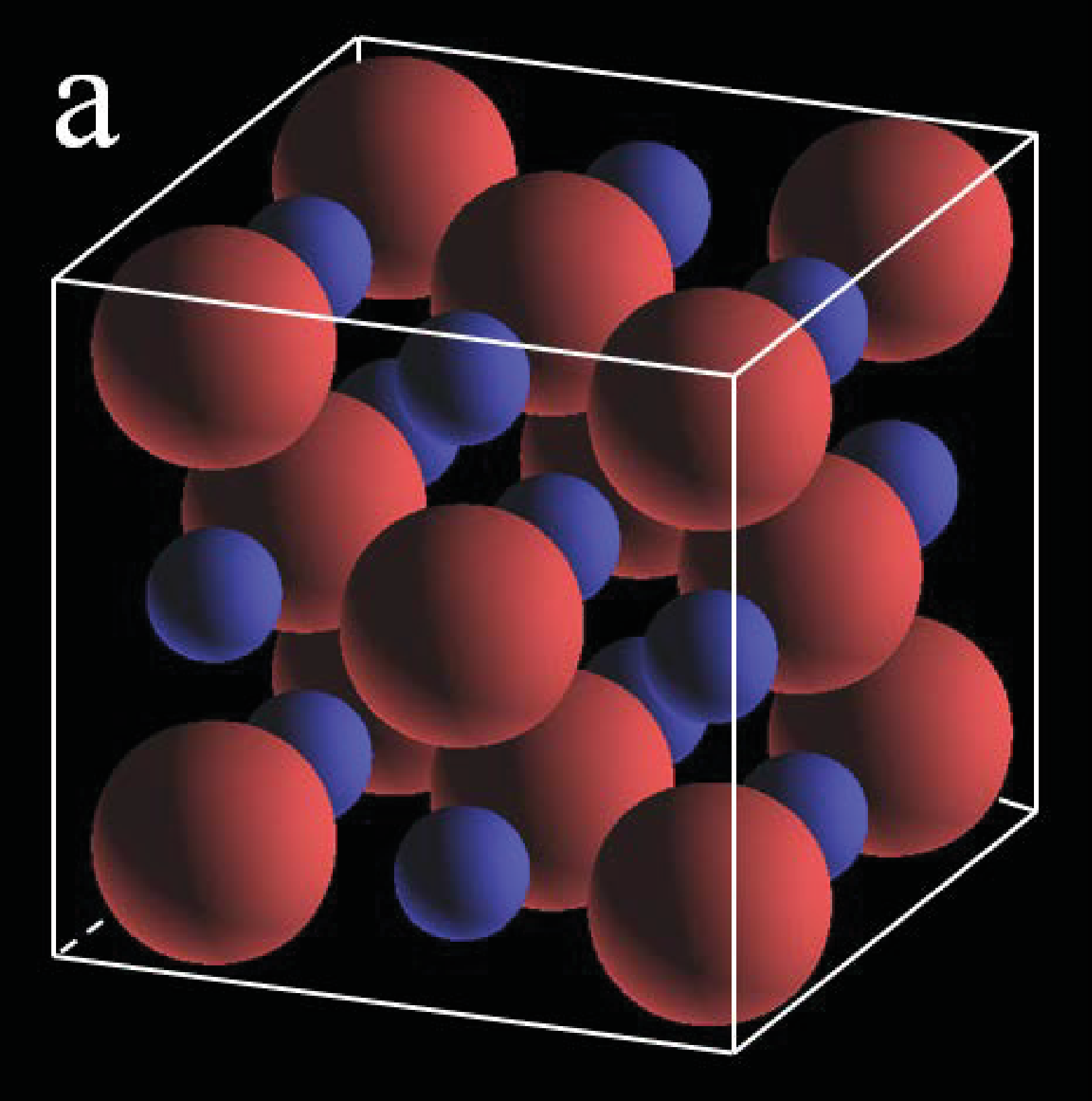}
\includegraphics[keepaspectratio,clip,height= 4.1cm,trim=0cm 0cm 0cm 0cm]{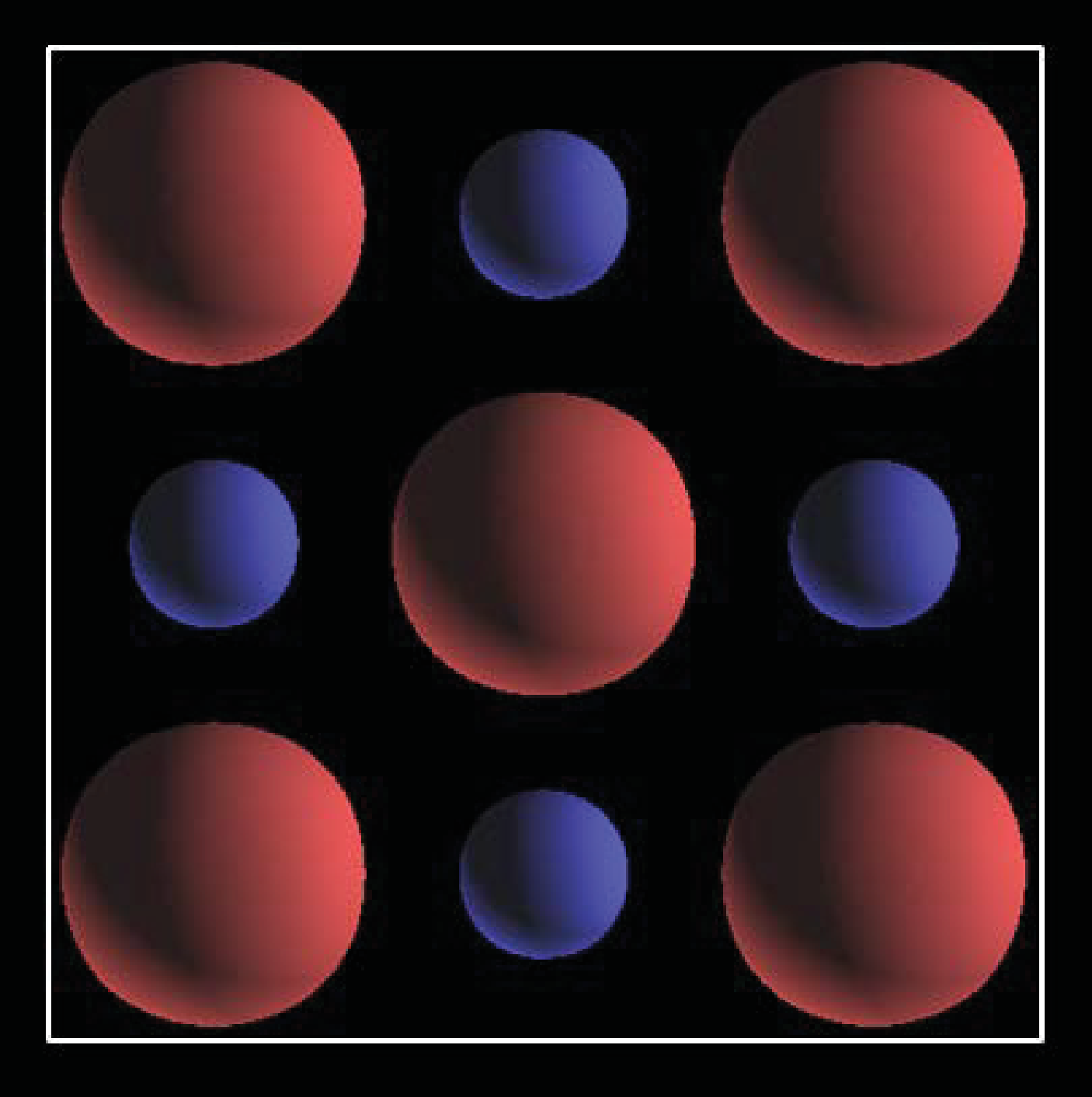}
\includegraphics[keepaspectratio,clip,height= 6.5cm,trim=0cm 0cm 0cm 0cm]{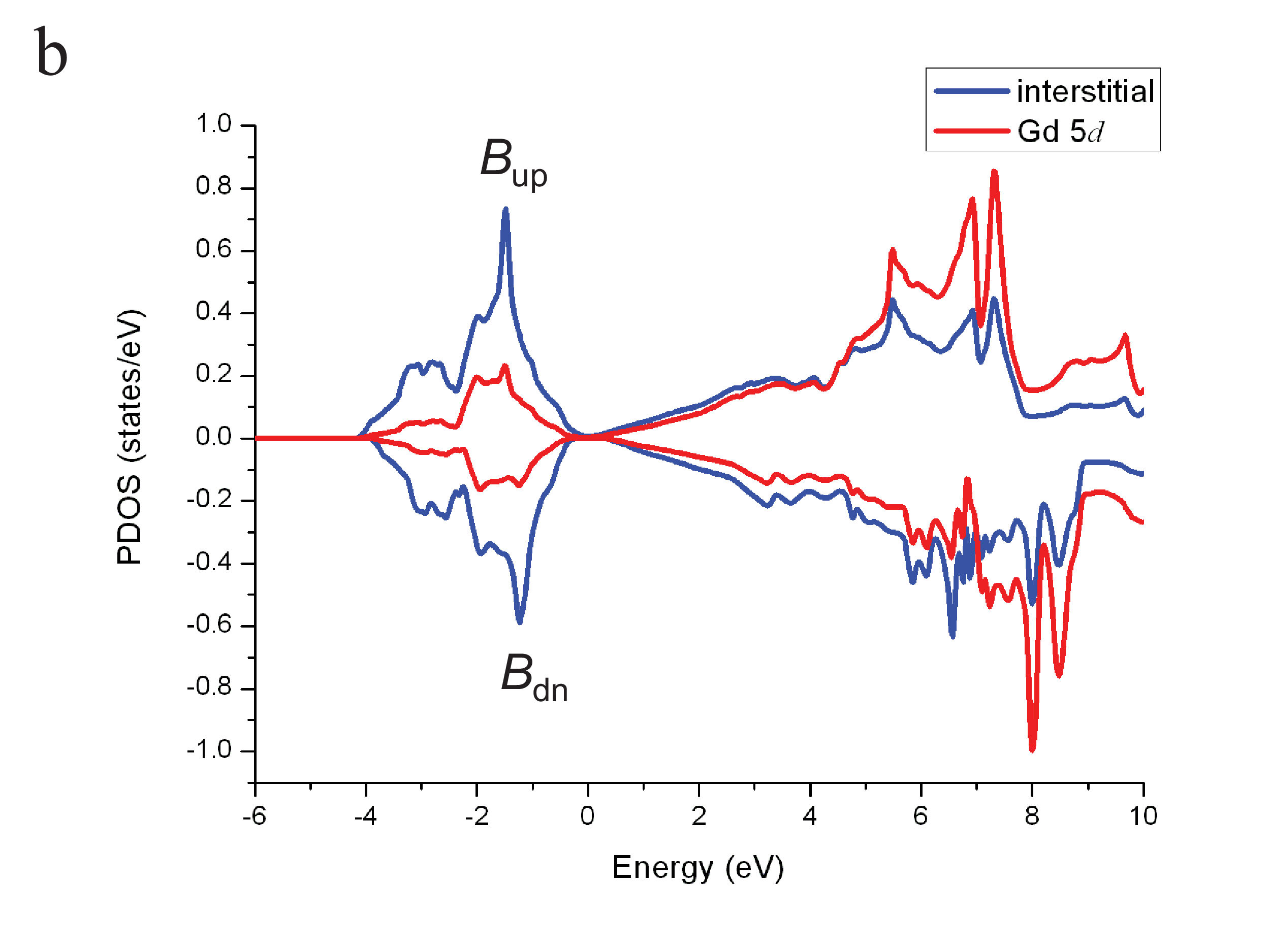}
\includegraphics[keepaspectratio,clip,height= 7cm,trim=0cm 0cm 6cm 0cm]{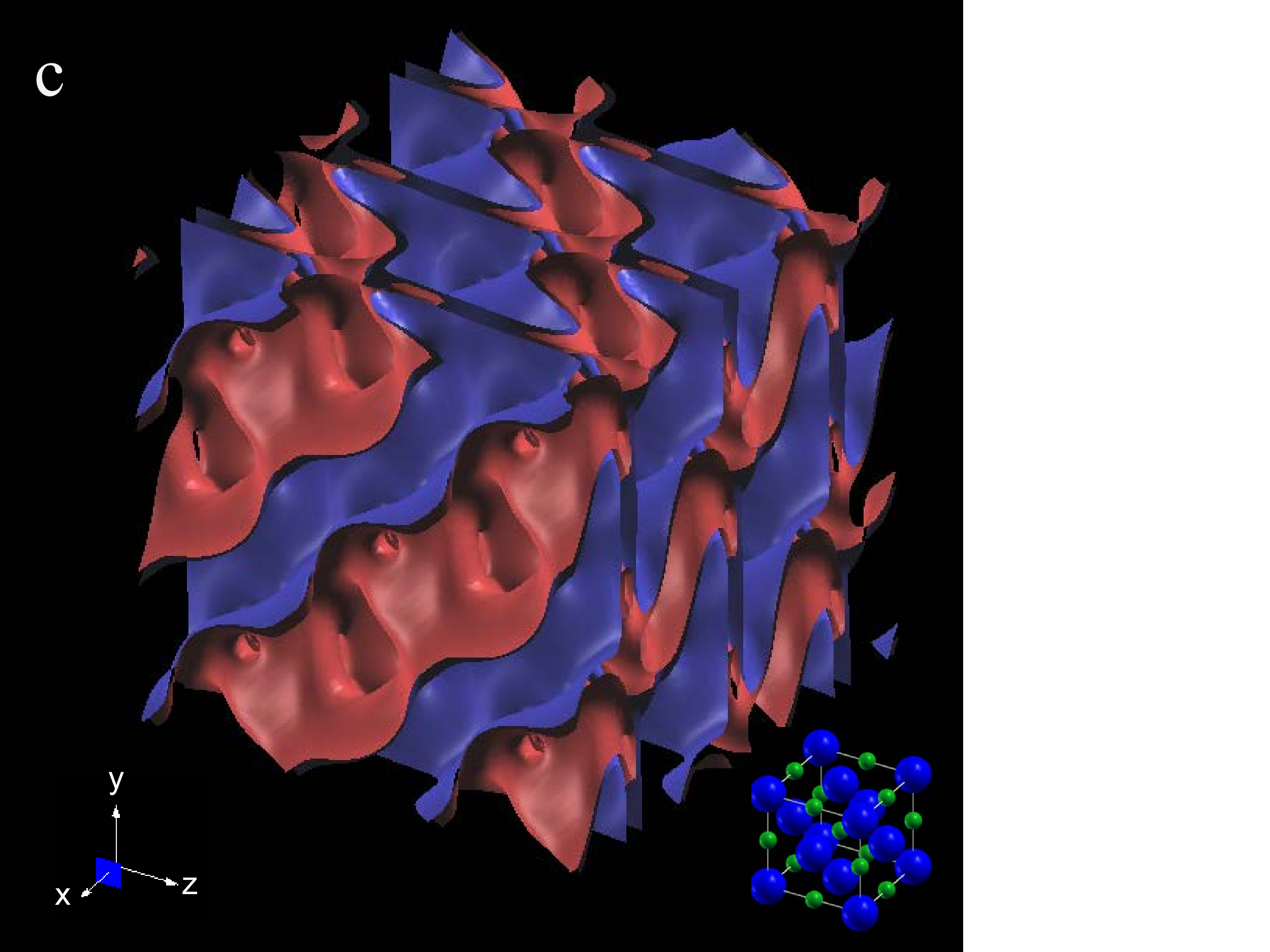}
\end{center}
\caption{\label{GdNspin}
({\bf a}) Calculated spin-density isosurfaces of a GdN bulk.
The right plot is viewed along the direction perpendicular to a cube face,
and has the isosurfaces of only the nearest atoms plotted for better visualization.
The plotted region represents a GdN conventional unit cell extended
by $30\%$ of the lattice constant, so that the spin isosurfaces of atoms
at face centers and corners can be entirely plotted.
Red stands for positive spin polarization, and blue for negative.
Gd and N atoms have nearly spherical positive and negative isosurfaces, respectively.
({\bf b}) Interstitial and Gd $5d$ PDOS of a ferromagnetic GdN bulk.
The main peaks of occupied majority- and minority-spin PDOS
are labeled as $B_{up}$ and $B_{dn}$, respectively.
({\bf c}) Isosurface of the spin probability density of the PDOS peak
at the value of $0.0002$\AA$^{-3}$.}
\end{figure}

We now consider bulk GdN.
A natural generalization of the spin bonding of the diagonal-dimer direct exchange
is to ask whether the ferromagnetic NN coupling in a GdN bulk is,
like the diagonal Gd dimer, also related to a spin bonding.
For this we plot the spin density of the GdN bulk in Fig.~\ref{GdNspin}a,
where it can be clearly seen that there is no obvious spin density lobe connecting two Gd atoms,
i.e., no spin bonding in a GdN bulk.
In fact, a previous DFT study of GdN \cite{GdN-dftprl}
already concludes that the ferromagnetism of the diagonal nearest neighbors
most likely originates from the RKKY interaction,
by analyzing the trends of exchange coupling of differently strained GdN compared to
all other gadolinium pnictides of larger-size anions.
RKKY is quite prominent in Gd bulk compounds, compared to
a pair of surface adatoms, because of the increased free-electron density
of states contributed by all the Gd. As we will show in the next paragraph,
the oscillating spin density in the interstitial region
makes direct exchange much less likely in the bulk.

To understand how Gd spins couple each other through
their delocalized electrons in a GdN bulk,
we plot both the interstitial and $5d$ PDOS.
This is shown in Fig.~\ref{GdNspin}b.
Within 5eV below the Fermi energy, the main PDOS peak around $E=-1.2$ eV
has a significant interstitial component but only a minor $5d$.
We further plot both the majority- and minority-spin orbitals associated with this peak,
shown in Fig.~\ref{GdNspin}c.
The isosurfaces of the orbitals spread all over the interstitial region.
This can be interpreted as the follows:
The Gd $5d$ electrons in the GdN bulk join the conduction electron sea,
and consequently they contribute to the RKKY interactions,
but do not form a directional orbital interaction
in between two Gd as in the surface dimer case.
Although the ferromagnetism seems to be understood quite well by RKKY,
the GdN-bulk study \cite{GdN-dftprl} does not exclude the possibility of
a ferromagnetic contribution from the $90^\circ$-superexchange interaction
which those authors note can be quite large in some oxides.
Indeed, from our finding of superexchange which becomes ferromagnetic
at an angle of 112 degrees, we predict that the 90-degree coupling in bulk
material has at least some ferromagnetic superexchange as well.

In summary, we have calculated the electronic structure of
coupled rare-earth (Gd) spins on a surface using
the PBE+U exchange correlation, in the first study of rare-earth spin-coupled adatoms.
The presence of Gd gives rise to rearrangement of the atomic structure
that is quite different from that of a Mn atom \cite{MnDFT}.
The geometry effect of the spin coupling is manifested by calculating
the exchange coupling between Gd atoms on the CuN surface and finding
antiferromagnetic coupling in a linear geometry and ferromagnetic in a diagonal geometry,
showing that the sign of $J$ can be tuned by different geometric arrangements.
We also find the Gd dimers in these two geometries have many similarities to
the nearest-neighbor (NN) and the next-NN Gd atoms in a GdN bulk.

The underlying physics of the dimers' magnetism is
studied by decomposing the magnetic couplings into the direct exchange,
superexchange, and RKKY interaction, and
the strength of the direct exchange is
further pictorially understood by the ``spin bonding'' between two Gd.
The diagonal case has ferromagnetic contributions from both direct
exchange and superexchange.
The antiferromagnetism for the linear geometry is
due to the predominance of superexchange for this configuration,
notwithstanding a large ferromagnetic direct exchange.
Superexchange is present in both geometries for the Gd dimer,
ferromagnetic for the diagonal and antiferromagnetic for the linear.
Even for the ferromagnetic coupling of the diagonal dimer,
superexchange constitutes 27\% of the total interaction.
While the bulk GdN compound is basically dominated by the RKKY interactions,
we find much smaller RKKY with the Gd dimers on a CuN surface.
Our calculations also show that the Gd spin of these structures is 7/2,
the same as that of GdN bulk, but different from a spin-4 free Gd atom with
a valance configuration $4f^{7}5d^{1}6s^2$.

\end{document}